\documentclass[aps,prd,reprint,nofootinbib,superscriptaddress]{revtex4-2}

\usepackage{amsmath,amssymb}

\usepackage{dsfont}
\usepackage{graphicx}
\usepackage{dcolumn}
\usepackage{bm}
\usepackage{physics}
\usepackage{mathtools}
\usepackage{tensor}
\usepackage{xcolor}
\usepackage{hyperref}
\usepackage{float}

\usepackage{multirow} 

\newcommand{\ii}{\mathrm{i}}
\newcommand{\ee}{\mathrm{e}}
\newcommand{\Id}{\mathds{1}}
\newcommand{\M}{\mathcal{M}}

\DeclareMathOperator{\sgn}{sgn}
\newcommand{\gbc}{\frac{\gamma^2-1}{\gamma^2+1}}
\newcommand{\pc}{\frac{2 \gamma}{\gamma^2+1}}

\newcommand{\s}{s}

\newcommand{\yad}{{y_{\mathrm{ad},\smallpm}}}

\newcommand{\yasy}{{y_{\mathrm{asy},\smallpm}}}

\newcommand{\smalltk}{\!\left(\text{\small $t,\vec{k}$}\right)}

\newcommand{\smallpm}{\text{\tiny{$\pm$}}}

\newcommand{\lcivitaT}{}
\def\lcivitaT#1#{\tensor#1{\varepsilon}}

\newcommand{\lcivita}{}
\def\lcivita#1#{\tensor#1{\epsilon	}}

\newcommand{\R}{}
\def\R#1#{\tensor#1{R}}

\newcommand{\J}{}
\def\J#1#{\tensor#1{J}}
\newcommand{\iJ}{}
\def\iJ#1#{\tensor#1{(J^{\scalebox{0.9}{-}{1}})}}

\newcommand{\F}{}
\def\F#1#{\tensor#1{F}}
\newcommand{\pF}{}
\def\pF#1#{\tensor#1{{F^{\text{\tiny (+)}}}}}
\newcommand{\mF}{}
\def\mF#1#{\tensor#1{{F^{\text{\tiny(--)}}}}}
\newcommand{\sF}{}
\def\sF#1#{\tensor#1{{{}^\star\! F }}}
\newcommand{\pmF}{}
\def\pmF#1#{\tensor#1{{F^{(\pm)\,} }}}
\newcommand{\sJ}{}
\def\sJ#1#{\tensor#1{{{}^\star\! J }}}
\newcommand{\pmJ}{}
\def\pmJ#1#{\tensor#1{{J^{(\pm)\,} }}}

\newcommand{\w}{}
\def\w#1#{\tensor#1{\omega}}
\newcommand{\pw}{}
\def\pw#1#{\tensor#1{{\omega^{\text{\tiny(+)}}}}}
\newcommand{\mw}{}
\def\mw#1#{\tensor#1{{\omega^{\text{\tiny(--)}}}}}
\newcommand{\pmw}{}
\def\pmw#1#{\tensor#1{{\omega^{\text{\tiny($\pm$)}}}}}
\newcommand{\sw}{}
\def\sw#1#{\tensor#1{{{}^\star\! \omega }}}

\newcommand{\ees}{}
\def\ees#1#{\tensor#1{{(e\wedge e)^{\star}}}}

\newcommand{\Pip}{}
\def\Pip#1#{\tensor#1{ {\Pi^{\text{\tiny (+)}}} }}
\newcommand{\Pim}{}
\def\Pim#1#{\tensor#1{ {\Pi^{\text{\tiny (--)}}} }}
\newcommand{\Pipm}{}
\def\Pipm#1#{\tensor#1{ {\Pi^{\text{\tiny ($\pm$)}}} }}

\newcommand{\Pig}{}
\def\Pig#1#{\tensor#1{ {\Pi^{\text{\tiny $(\gamma)$}}} }}
\newcommand{\Hodge}{}
\def\Hodge#1#{\tensor#1{ {\star} }}
\newcommand{\sR}{}
\def\sR#1#{\tensor#1{ {{}^{\star}\! R\,} }}%
\newcommand{\Rs}{}
\def\Rs#1#{\tensor#1{ {R^{\star}\,}}}%
\newcommand{\ssR}{}
\def\ssR#1#{\tensor#1{ {{}^{\star\star}\! R\,} }}%
\newcommand{\Rss}{}
\def\Rss#1#{\tensor#1{ {R^{\star\star}\,}}}%
\newcommand{\pR}{}
\def\pR#1#{\tensor#1{ {{}^{\text{\tiny (+)}}\! R\,} }}
\newcommand{\Rp}{}
\def\Rp#1#{\tensor#1{ {R^{\text{\tiny (+)}}\,}}}
\newcommand{\mR}{}
\def\mR#1#{\tensor#1{ {{}^{\text{\tiny (--)}}\! R\,} }}
\newcommand{\Rm}{}
\def\Rm#1#{\tensor#1{ {R^{\text{\tiny (--)}}\,}}}
\newcommand{\pmR}{}
\def\pmR#1#{\tensor#1{ {{}^{\text{\tiny ($\pm$)}}\! R\,} }}
\newcommand{\Rpm}{}
\def\Rpm#1#{\tensor#1{ {R^{\text{\tiny ($\pm$)}}\,}}}
\newcommand{\mpR}{}
\def\mpR#1#{\tensor#1{ {{}^{\text{\tiny ($\mp$)}}\! R\,} }}
\newcommand{\Rmp}{}
\def\Rmp#1#{\tensor#1{ {R^{\text{\tiny ($\mp$)}}\,}}}
\newcommand{\B}{}
\def\B#1#{\tensor#1{ B }}

\newcommand{\dualarrow}{}
\def\dualarrow#1#{\tensor#1{{\mapsto
}}}
\newcommand{\DualRot}{}
\def\DualRot#1#{\tensor#1{{\mathfrak{D}}}}

\newcommand{\comment}[1]{#1}

\begin{document}

\title{Spinfoams, \boldmath{$\gamma$}-duality and parity violation in primordial gravitational waves}
\author{Eugenio Bianchi}
\email[]{ebianchi@psu.edu}
\author{Monica Rincon-Ramirez}
\email[]{mramirez@psu.edu}

\affiliation{Institute for Gravitation and the Cosmos, The Pennsylvania State University, University Park, Pennsylvania 16802, USA}
\affiliation{Department of Physics, The Pennsylvania State University, University Park, Pennsylvania 16802, USA}

\begin{abstract}
The Barbero-Immirzi parameter $\gamma$ appears as a coupling constant in the spinfoam dynamics of loop quantum gravity. In this work, we highlight that $\gamma$ can be understood as a measure of gravitational parity violation via a duality rotation for the EPRL spinfoam model. We call this property $\gamma$-duality, and we investigate an effective field theory for gravity and a scalar field with the same degree of parity violation. The resulting relation between the coupling constants of parity-even and parity-odd higher-curvature terms in the effective action is determined by $\gamma$, opening the possibility of its measurement in the semiclassical regime. For a choice of $\gamma$-dual effective action, we study cosmic inflation and show that the observation of a primordial tensor polarization, together with the tensor tilt and the tensor-to-scalar ratio, provides a measurement of the Barbero-Immirzi parameter and, therefore, of the scale of discreteness of the quantum geometry of space.
\end{abstract}

\maketitle

\section{Introduction}
The Barbero-Immirzi parameter $\gamma$ plays a central role in Loop Quantum Gravity (LQG)  \cite{Rovelli:2014ssa,Thiemann:2007pyv,Ashtekar:2021kfp}. The parameter sets the typical scale of the spectrum of geometric observables such as the volume of a quantum of space, the area of the interface between two quanta and the length of a curvature defect line \cite{Rovelli:1994ge,Ashtekar:1996eg,Ashtekar:1997fb,Bianchi:2012wb,Bianchi:2008es}. The elementary area operator has eigenvalues
\begin{equation}\label{area spectrum}
    A_j=8\pi G\hbar\, \gamma\, \sqrt{j(j+1)}\,, \qquad j \in \left\{0,\,\tfrac{1}{2},\,1,\,\tfrac{3}{2},\ldots\right\}\,.
\end{equation}
In particular, the theory predicts an area gap $a_*=4\sqrt{3}\,\pi\,\gamma\, \ell_P^2\,$, where $\ell_P=\sqrt{G\hbar/c^3}\simeq 1.6\times 10^{-35}\,\mathrm{m}$ is the Planck length and the Barbero-Immirzi parameter $\gamma>0$ is a dimensionless coupling constant that, in principle, is to be determined experimentally.
At the classical level \cite{BarberoG:1994eia,Immirzi:1996di,Holst:1995pc,Hojman:1980kv}, the parameter $\gamma$ can be understood as a coupling constant in the Einstein-Cartan-Holst action for gravity. This action is invariant under orientation-preserving diffeomorphisms but not under orientation-reversing ones, and therefore is parity violating:
\begin{equation}\label{eq:S-ECH}
S=\frac{1}{16 \pi G}\!\int \frac{1}{2}\epsilon_{IJKL}\,e^I\wedge e^J\wedge F^{KL}- \frac{1}{\gamma}\, e_I\wedge e_J\wedge F^{IJ}\,,
\end{equation}
where the first term in the integral is the parity-odd Einstein-Cartan density and the second is the parity-even Holst density. At the quantum level, spinfoams provide a non-perturbative definition of the covariant dynamics of LQG \cite{Perez:2012wv}. The spinfoam dynamics is not defined in terms of an action for fields on a manifold, but instead by a spinfoam vertex amplitude. The Engle-Pereira-Rovelli-Livine (EPRL) \cite{Engle:2007wy}  vertex amplitude $W_\gamma$ depends on the Barbero-Immirzi parameter $\gamma$ and belongs to a one-parameter family of spinfoam models related by a duality rotation with an angle $\theta$. In the limit $\theta\to 0$ at fixed eigenvalues of the area, the EPRL model reduces to the Barrett-Crane model (BC) \cite{Barrett:1999qw},  which is parity invariant. On the other hand, the EPRL model with finite value of $\gamma$ is parity-violating \cite{RovelliW-EDiscrete} and defined by
\begin{equation}\label{eq:theta-gamma}
\tan\theta=\frac{1}{\gamma}\,.
\end{equation}
\comment{Having identified this exact property of the EPRL spinfoam model, 
which we call $\gamma$-duality, motivates us to investigate the formulation of an effective action for gravity with the same amount of parity violation determined by the duality rotation \eqref{eq:theta-gamma}. This requirement assumes that $\gamma$-duality is exactly realized at the semiclassical level and not broken by other effects, which would have to be studied in conjunction with the effective action studied in this work. 

At the semiclassical level, we consider the one-parameter family of theories with a  duality rotation of the curvature}
\begin{align}
F^{IJ}\quad&\overset{\theta}{\longrightarrow}\quad+\cos\theta\;\, F{^{IJ}}\;+\;\sin\theta\;\; {}^*\!F^{IJ}\,,\label{eq:duality-rotation-def}\\[.5em]
{}^*\!F^{IJ}\quad&\overset{\theta}{\longrightarrow}\quad-\sin\theta\;\: F^{IJ}\;+\;\cos\theta\;\; {}^*\!F^{IJ}\,,\nonumber
\end{align}
where ${}^*\!F_{IJ}=\frac{1}{2}\epsilon_{IJKL}\,F^{KL}$ is the Hodge dual in the internal Lorentz indices. Starting from an effective action for gravity that is parity non-violating, we can produce a parity-violating effective action via a duality rotation of the curvature. For instance, the action \eqref{eq:S-ECH} can be obtained by a duality rotation of the Einstein-Cartan density with angle $\theta$ satisfying the condition \eqref{eq:theta-gamma}. We note that the duality rotation is not a symmetry of the action, or of its solutions \cite{Kol:2022bsd,Kol:2023yxd,Monteiro:2023dev}, but a relation between theories with different values of their coupling constants. 

When the effective action is organized as a derivative expansion \cite{Weinberg}, each term in the expansion is a Lagrangian density that comes with a coupling constant which, in principle, needs to be fixed by observations. Remarkably, the requirement that the effective action is obtained by a duality rotation of a parity non-violating effective action introduces a non-trivial relation between these coupling constants --- a single parameter $\gamma$ sets the scale of all parity violating terms in the action, modeling at the semiclassical level an exact property of the spinfoam dynamics $W_\gamma$. An effective field theory that satisfies this requirement is described by an action $S_\gamma$ that we call $\gamma$-dual.

\medskip

While a top-down derivation of the effective action directly from the non-perturbative spinfoam dynamics $W_\gamma$ is still missing, the condition of $\gamma$-duality allows us to relate coupling constants of parity-even and parity-odd terms in the effective action $S_\gamma$. Therefore the observation of gravitational parity violation would provide an indirect measurement of the Barbero-Immirzi parameter $\gamma$, and therefore of the LQG area gap. In this paper we investigate a $\gamma$-dual model of cosmological inflation and show how the spectrum of primordial gravitational waves depends on the coupling constant $\gamma$.

\medskip

The effect of gravitational parity violation in primordial inflation has been studied for the first time in \cite{LueWangKamionkowski} where a Chern-Simons modification of General Relativity was considered and a non-vanishing polarization $\Pi\equiv(A_{T+}-A_{T-})/(A_{T+}+A_{T-})$ is shown to arise for the amplitude $A_{T\pm}$ of primordial tensor modes with circular polarization $(\pm)$. We refer to \cite{AlexanderYunes} for a review and to \cite{JackwiPi,Alexander:2004us,Alexander:2004wk,Lyth:2005jf,ContaldiMagueijoSmolin,SatohSoda,Satoh2010,Dyda:2012rj,Kawai:2017kqt,Gialamas:2022xtt,Bartolo:2014hwa,Bartolo:2015dga,Bartolo:2017szm,Nojiri:2019nar,Bordin:2020eui,Alexander:2022cow,Gong:2023kpe,Jenks:2023pmk,Creque-Sarbinowski:2023wmb} for older and more recent developments. Our analysis is motivated by a relation between parity violation in primordial gravitational waves and the Barbero-Immirzi parameter in LQG originally proposed in \cite{ContaldiMagueijoSmolin}. We develop this proposal by introducing a new crucial ingredient --- spinfoam $\gamma$-duality. In general, at the quadratic order in curvature, the Lagrangian for gravity and an inflaton scalar field $\phi$ can include both a parity-even term  $L_{GB}$ proportional to the  Gauss-Bonnet density and a parity-odd term $L_{CS}$ proportional to the Pontryagin density (also know as Chern-Simons term) \cite{SatohSoda, Satoh2010}, with
\comment{
\begin{align}
L_{GB}=&\,f_{GB}(\phi)\; \tfrac{1}{2}(R_{\mu\nu\rho\sigma}R^{\mu\nu\rho\sigma}\!-\!4R_{\mu\nu}R^{\mu\nu}\!+\!R^2)\,,\label{L_GB}\\[.5em]
L_{CS}=&\,f_{CS}(\phi) \;\tfrac{1}{4}\tfrac{\epsilon^{\mu\nu\rho\sigma}}{\sqrt{-g}}\,R^\alpha{}_{\beta\mu\nu}\,R^\beta{}_{\alpha\rho\sigma}\,.\label{L_CS}
\end{align}
The two functions $f_{GB}(\phi)$ and $f_{CS}(\phi)$ are in principle independent. The requirement of $\gamma$-duality of the action imposes a relation between the two functions: $2f_{GB}(\phi)/f_{CS}(\phi)=\gamma-\frac{1}{\gamma}$.} In this paper we show that, in a $\gamma$-dual model of slow-roll inflation, the asymmetry $\Pi$ of the amplitude of circular polarizations, the tensor-to-scalar ratio $r\equiv (A_{T+}+A_{T-})/A_S$ and the tensor tilt $n_T$ are related to the parameter $\gamma$ by the equation
\begin{equation}\label{gamma-r}
\frac{1}{\gamma}-\gamma=\frac{\pi}{8}\,\frac{r+8\,n_T}{\Pi}.
\end{equation}
Therefore the observation of a primordial parity violation $\Pi\neq 0$ and a violation of the $(r,n_T)$ `consistency relations', $r\neq-8\,n_T$, would provide a measurement of the Barbero-Immirzi parameter $\gamma$, and thus fix the scale $a_*$ of the area gap.

\medskip

The analysis presented here is complementary to the framework of loop quantum cosmology \cite{MB:SingularityLQC, AA-MB-JL:LQC-Math,AA-PS:LQC, IA-PS:LQC, IA-EWEea} where Planck scale effects on the background dynamics and on metric perturbations during the pre-inflationary era are found to leave phenomenological imprints in cosmic microwave background observables. The analysis is also complementary to the approach of spinfoam cosmology \cite{Bianchi:2010zs,Gozzini:2019nbo,Frisoni:2023lvb} where one investigates the cosmological regime of the full spinfoam dynamics. The phenomenological effects considered in this paper have their origin in the inflationary era, where the curvature is already far from the Planck scale and an effective field theory motivated by spinfoams can be used. In this sense, the analysis follows the same logic used in the study of parity violation in the coupling of fermions to gravity and torsion in an effective field theory \cite{Perez:2005pm,Freidel:2005sn}, with the additional new ingredient of $\gamma$-duality introduced here.

Specifically, we adopt the point of view that $\gamma$ is a coupling constant that needs to be determined experimentally, as any other coupling constant. This viewpoint is adopted also in recent analysis in loop quantum cosmology \cite{Ashtekar:2015iza,Agullo:2016hap}, where observational bounds on the area gap $a_*$ --- and therefore on $\gamma$ --- are investigated.  This point of view differs from the one (adopted for instance in \cite{IA-PS:LQC}) where  $\gamma$ is set to a fixed exact value $\gamma_0=\log(2)/\pi\sqrt{3} $ \cite{AABaezCorichi} or $\gamma_M$  \cite{Meissner2004}, which is obtained by imposing that the microscopic entropy of the horizon-area ensemble matches the thermodynamics entropy of a black hole of the same horizon area. Other analysis of the derivation of the Bekenstein-Hawking entropy formula show that the coefficient of the area law is determined by semiclassical infrared phenomena and is independent of the value of the Barbero-Immirzi parameter  \cite{Jacobson:2007uj,Bianchi:2012ui,Ghosh:2012wq,BarberoG:2022ixy}, and  therefore needs to be determined experimentally.

\medskip

The paper is organized as follows. In section \ref{sec: gDuality Spinfoams} we describe $\gamma$-duality in spinfoams, both at the level of the non-perturbative definition of the vertex amplitude and at the semiclassical level. In section \ref{sec: Orientations and Diff-} we discuss the link between parity, orientation-reversing diffeomorphisms and internal Lorentz reversals, and describe how different gravitational terms in the effective action transform. In section \ref{sec:Model} we use duality rotations to  construct a spinfoam-motivated effective action for inflation and discuss the assumptions of the model. In section \ref{sec: Inflation and PGW} we compute the primordial power spectrum of primordial gravitational waves, and in section \ref{Section: Discussion measurements} we discuss prospects for observations.

\section{\boldmath{$\gamma$}-Duality in Spinfoams}\label{sec: gDuality Spinfoams}
We briefly discuss the EPRL spinfoam model \cite{Rovelli:2014ssa,Perez:2012wv}, illustrating its parity-violating nature \cite{RovelliW-EDiscrete,Bianchi:2009ri,Bianchi:2011hp,Neiman:2011gf}  and the role of $\gamma$-duality both in its non-perturbative definition and  in the semiclassical limit.

\medskip

The Barbero-Immirzi parameter appears as a dimensionless coupling constant in the Einstein-Cartan-Holst action $S_{ECH}$, \eqref{eq:S-ECH}. This action can be obtained starting from the Einstein-Cartan action $S_{EC}$,
\begin{align}
&S_{EC} =\frac{1}{2\kappa}\,\int \frac{1}{2}\epsilon_{IJKL}\,e^I\wedge e^J\wedge F^{KL}\label{eq:S-EC}\\[.5em]
&\overset{\theta}{\longrightarrow}
\quad
S_{ECH}=\frac{\cos\theta}{2\kappa}\,\int \frac{1}{2}\epsilon_{IJKL}\,e^I\wedge e^J\wedge F^{KL}\nonumber\\[.5em]
&\hspace{10.2em} -\tan\theta\; e_I\wedge e_J\wedge F^{IJ}\label{ECH1}\,,
\end{align}
via the duality rotation \eqref{eq:duality-rotation-def} of the curvature. Comparing the expressions \eqref{eq:S-ECH} and \eqref{ECH1}, we observe that  the Barbero-Immirzi parameter is related to the duality rotation angle $\theta$ by the relation \eqref{eq:theta-gamma} and $\kappa$ is related to the observed value of Newton's constant $G$ by $8\pi G=\kappa/\cos\theta$. 

\bigskip

In spinfoam quantum gravity one finds a similar relation at the non-perturbative level. The relation can be expressed in terms of the labels $(p,j)$ of the unitary irreducible representations of the principal series of the Lorentz group $SL(2,\mathbb{C})$,
\begin{equation}
g\in SL(2,\mathbb{C}) \qquad\longrightarrow \quad D^{(p,j)}(g)\,.
\end{equation}
The Engle-Pereira-Rovelli-Livine spinfoam model (EPRL) with Barbero-Immirzi parameter $\gamma$ \cite{Engle:2007wy} is defined by the representation  $D^{(\gamma j,j)}(g)$, while the Barrett-Crane model (BC) \cite{Barrett:1999qw} by the representation $D^{(p_0,0)}(g)$. The two models are related by the duality rotation
\begin{equation}\label{eq:SpinfoamsPJtheta}
p\;=\;\cos\theta \;p_0\;,\qquad j\;=\;\sin\theta\;p_0\;,
\end{equation}
where the angle $\theta$ satisfies \eqref{eq:theta-gamma}, i.e.,
\begin{equation}\label{eq:SpinfoamsPJgamma}
p\,=\,\gamma\,j\,.
\end{equation}
The BC model is obtained from the EPRL model as the limit $\theta\to 0$ at fixed $p_0$. Equivalently, this duality rotation  defines a one-parameter family of parity-violating EPRL models starting from the  parity-non-violating BC model as a seed. 

\comment{Specifically, the building block of a spinfoam model is the wedge amplitude $\mathcal{A}_w$ which encodes the Lorentz parallel transport from a source edge to a target edge of a spinfoam vertex, forming a wedge. We refer to \cite{RovelliVidotto} for a standard introduction to the techniques used to define the wedge amplitude.} The spinfoam wedge amplitude $\mathcal{A}_w$ for a coherent boundary state labeled by the spinors $\zeta$ and $\zeta'$ takes the form \cite{Barrett:2009mw}
\begin{equation}
\label{eq:wedge-amplitude}
\mathcal{A}_w(g,\zeta',\zeta'')=[\,j_w,\zeta'|D^{(p_w,j_w)}(g)|j_w,\zeta''\rangle=\!\!\int_{\mathbb{C}P^1}\!\!\!d\mu\;\;\ee^{\,\ii S_w},
\end{equation}
where the wedge action $S_w$ consists of the sum of two terms proportional to the quantum numbers $p_w$ and $j_w$ that label the representation, i.e., $S_w= p_w\, \Theta_w\;+\;j_w \,\chi_w$. In the semiclassical limit for a Lorentzian $4$-simplex, a saddle-point analysis shows that, for $p_0\to\infty$, the integral is dominated by the exponential of the sum of the wedge actions evaluated at each saddle point,
\begin{equation}
S_v=\sum_w \bar{S}_w\;=\;\sum_w p_w\, \bar{\Theta}_w\;+\;\sum_w j_w \,\bar{\chi}_w\,.
\label{eq:S-w}
\end{equation}
The two terms appearing in the semiclassical action for a spinfoam vertex can be understood as a discretization on a Lorentzian $4$-simplex of the two terms in the Einstein-Cartan-Holst action \eqref{ECH1}. The first of the two terms reproduces the Regge action for a Lorentzian $4$-simplex \cite{Regge:1961px},
\begin{equation}
\frac{1}{8\pi G \hbar}S_{\mathrm{Regge}}\;=\;\sum_w \frac{A_w}{8\pi G \hbar} \bar{\Theta}_w\,,
\end{equation}
where $A_w$ is the area of the triangle where the wedge $w$ hinges, and $\bar{\Theta}_w$ is the boost angle between the two space-like tetrahedra that define the wedge. This relation is consistent with the area spectrum in loop quantum gravity in the large spin limit,
\begin{equation}
A_w=8\pi G\hbar\; p_w\;=\;8\pi G\hbar\,\gamma\, j_w\,.
\end{equation}
On the other hand, the second term in \eqref{eq:S-w} includes a twist angle $\bar{\chi}_w$ that depends on the phase of the coherent boundary state spinors and does not affect the dynamics of the model. This term is analogous to the Holst term in \eqref{ECH1} and has the same transformation properties under parity\footnote{Their parity can be read for instance in the asymptotics of the vertex amplitude for Lorentzian boundary data, where a sum of saddle points results in the two terms of different parity $A_v\sim \ee^{+\ii \sum_w p_w\, \bar{\Theta}_w\;+\;\ii \sum_w j_w \,\bar{\chi}_w}+\ee^{-\ii \sum_w p_w\, \bar{\Theta}_w\;+\;\ii\sum_w j_w\,\bar{\chi}_w}$.}. Note also that the Barrett-Crane model is obtained as the limit $\theta\to 0$ where this second term is not present. When changing the angle $\theta$, the geometric relation between the area $A_w$ and the representation $p_w$ can be preserved by rescaling $\kappa$ by $\cos \theta$ as discussed after \eqref{ECH1} at the level of the classical action.\\

The argument described above shows that the duality rotation \eqref{eq:duality-rotation-def} relates the semiclassical limit of spinfoam models with different values of the Barbero-Immirzi parameter. Remarkably, this is an exact relation that holds beyond the semiclassical limit $p_0\gg 1$. The defining equations of the EPRL spinfoam model with Barbero-Immirzi parameter $\gamma$ is the linear simplicity constraint \cite{Rovelli:2014ssa}
\begin{equation}
\vec{K}=\gamma\vec{L}
\end{equation}
where $\vec{K}$ and $\vec{L}$ are the generators of boosts and rotations for the Lorentz group.\footnote{Specifically, let us denote $J^{IJ}$ the generators of the Lorentz group $SL(2,\mathbb{C})$ and introducing an orthonormal frame $e^I_\mu=(u^I,\xi_i^I)$ with $\eta_{IJ} u^I u^J=-1$. We can then define the generators of the $SU(2)$ rotations group $L_i=\frac{1}{2}\epsilon_{IJKL}J^{IJ}u^K\xi^L_i$ that preserve the timelike direction $u^I$, and the generators of boosts $K_i=J_{IJ}\xi^I_i u^J$. } In terms of the self-dual and the anti-self-dual generators,
\begin{equation}
J_i^{(\pm)}=\frac{L_i\pm\ii\, K_i}{2}\,,
\end{equation}
the linear simplicity constraint takes the form
\begin{equation}
J_i^{(\pm)}=\frac{1\pm\ii\,\gamma}{\sqrt{1+\gamma^2}}\,J^{(0)}_i\,
\end{equation}
where $J^{(0)}_i$ is defined by the parity-even limit $\gamma\to\infty$. We can now introduce a duality rotation \cite{Speziale:2012nu}
\begin{equation}
J_i^{(\pm)}\;\overset{\theta}{\longrightarrow} \;\;\ee^{\pm\ii \theta}\, J_i^{(\pm)}\,,
\end{equation}
and notice again that the EPRL and the BC models are consistently related by a duality rotation with angle $\theta$ given by \eqref{eq:theta-gamma}. In a (formal) connection representation $\Psi[\omega_i^{(+)},\omega_i^{(-)}]$ where the self-dual/anti-selfdual generators act as a derivative, $J_i^{(\pm)}\Psi[\omega_i^{(+)},\omega_i^{(-)}]=-\ii\frac{\delta}{\delta \omega_i^{\pm}}\Psi[\omega_i^{(+)},\omega_i^{(-)}]$, the duality transformation relates the amplitude of states for theories with different value of the Barbero-Immirzi parameter $\gamma$, with the requirement that, in the limit $\gamma\to\infty$ at fixed elementary areas, the theory is parity invariant. Any spinfoam vertex that satisfies this condition, and therefore implements the relation \eqref{ECH1} at the non-perturbative level, is called $\gamma$-dual. The EPRL model is a specific and concrete proposal in this class. In the rest of this paper we take this property, $\gamma$-\emph{duality}, as fundamental and we explore its consequences in a semiclassical limit described by an effective action.

\section{Parity and Orientation-Reversing Diffeomorphisms}\label{sec: Orientations and Diff-}

In this section we discuss the notion of parity in a general relativistic theory where geometry is dynamical. In particular we show how orientation-reversing diffeomorphisms generalize the notion of parity defined in a special relativistic theory, and clarify the link between orientation-reversing diffeomorphisms and orientation reversals in the internal Lorentz group in the Einstein-Cartan formalism.  We then describe how different gravitational terms in the effective action transform.\\

In 3-dimensional Euclidean space, a \textit{parity transformation} $P\,:\,\lbrace x,y,z\rbrace\mapsto \lbrace -x,-y,-z\rbrace$ is defined as the simultaneous reversal of the three spatial Cartesian axes, that is the product of the three spatial-inversion operators with, e.g., $P_x:\,x\mapsto -x$. Similarly, a time reversal $T: \,t\mapsto -t$ is a flip of the temporal axis. In a special relativistic theory, these specific transformations clearly depend on a choice of an inertial rest frame in Minkowski space. It is then useful to think of them in terms of Lorentz transformations that are not connected to the identity. The Lorentz group $O(1,3)$ is the group of transformations $\Lambda^I{}_J$ that preserve the Minkowski metric $\eta_{IJ}$. The parameter space of Lorentz transformations splits into four connected components $O(1,3)\simeq L_+^\uparrow \cup L_+^\downarrow\cup L_-^\uparrow \cup L_-^\downarrow$, with the subgroup of proper orthochronous Lorentz transformations $O_0(1,3)=SO^\uparrow(1,3)\simeq L_+^\uparrow$ being the components connected to the identity \cite{PeskinSchroeder,streater2016pct}. We can then consider Lorentz transformations modulo transformations connected to the identity,
\begin{equation}
O(1,3)/O_0(1,3)\;=\;\{\Id,\,P,\,T,\,PT\}\,,
\end{equation}
with the quotient space generated by parity $P$ and time reversal $T$. Orientation reversals of Minkowski space are Lorentz transformations $\Lambda$ with parameters in $L_+^\downarrow\cup L_-^\uparrow\,=\, T  L_+^\uparrow\cup  P  L_+^\uparrow$ and are defined by $\det(\Lambda)=-1$.

\medskip

The notions of parity and time reversal described above require a fixed background metric such as the Minkowski metric. Here we are interested in gravitational parity violation, where the metric and its causal structure are not fixed as a background structure but are dynamical instead. Before defining orientation reversals for a spacetime manifold, let us start from a vector space, a linear space of dimension $d$ that is not equipped with any additional structure. In this case, given an \textit{ordered} basis of $d$ linearly independent vectors $E_I=\{E_1,\ldots,E_d\}$, we can consider a change of basis given by a matrix $M$. The requirement that  $\tilde{E}_{I} \, =\, M_I{}^{J}\,E_J$ is still a linear basis imposes that $\det M \neq 0$. Therefore we have two distinct equivalence classes of bases:  If $\det(M)>0$ then the two bases belong to the same class; if $\det(M)<0$, they belong to different classes. Each one of such equivalence classes is an \textit{orientation} $\Xi$ for the vector space.

In the case of a differentiable $d$-dimensional manifold $\mathcal{M}$, an orientation $\Xi$ can be equivalently defined as either \cite{Tu, BottTu, BaezMuniain}: (i) A continuous point-wise orientation on the manifold; (ii) An equivalence class of \textit{nowhere-vanishing} \textit{top-forms} (or $d$-forms in a $d$-dimensional manifold); or, (iii) An equivalence class of oriented atlases. A manifold is called orientable if it is possible to make a choice of orientation. If $\M$ is connected and orientable, there are again only two possible orientations $\Xi=\pm1$.

Definition (i) is the generalization of the concept of orientation of a vector space introduced above, where one considers a frame field $E_I$ (i.e., a set of bases $E_I(p)$, one at each point $p\in\M$), with the additional condition that the frame must have a point-wise continuous class: at every point $p$ there is a neighborhood with the same choice of orientation $\Xi_p$ of its tangent space.\footnote{We note that this definition of orientation of a manifold does not require the existence of a global continuous basis. The frame can be discontinuous, as long as the point-wise orientation is continuous.}

Definition (ii) is the one that is of the most relevance here. Let us discuss briefly its equivalence to (i). Considering a \textit{nowhere-vanishing} top-form $\Omega$, and an orientation $\Xi=[E_I]$ for which $E_I$ is a representing frame, we can check if the inner product $\langle \Omega, E_1\cdots E_d\rangle$ is either positive or negative as, by definition, it cannot be zero. This defines an equivalence relation for nowhere-vanishing top-forms, splitting them into two disjoint sets. The orientation $\Xi$ of the manifold $\mathcal{M}$ is given by an equivalence class of smooth nowhere-vanishing top-forms $\Omega$ and we have $\Xi=[E_I]=[\Omega]$ with $\langle \Omega, E_1\cdots E_d\rangle>0$.

Definition (iii) in terms of an oriented atlas is specifically useful once we introduce coordinates $x^\mu$ and dynamical fields over the manifold, as done in General Relativity. We set $d=4$ and assume that the $4$-dimensional manifold $\mathcal{M}$ representing spacetime is orientable.  The orientation $\Xi=[\partial_\mu]$ in each chart of the atlas is given by the coordinate frame $\partial_\mu=\partial/\partial x^\mu$. A spacetime diffeomorphism $\varphi\in \mathrm{Diff}(\mathcal{M})$ is then a smooth function $y^\mu=\varphi^\mu(x)$ with smooth inverse. Under a spacetime diffeomorphism, the nowhere vanishing $4$-form $dx^0\wedge\cdots\wedge dx^3$ transforms as
\begin{equation}
dy^0\wedge dy^1\wedge dy^2\wedge dy^3
\,=\,\det(J^{\mu}{}_\nu) \,dx^0\wedge dx^1\wedge dx^2\wedge dx^3\,,\nonumber
\end{equation}
where $J^\mu{}_\nu=\partial y^\mu/\partial x^\nu$ is the Jacobian of the transformation. An \emph{orientation-preserving} diffeomorphism is characterized by $\det(J)>0$, an \emph{orientation-reversing} diffeomorphism by $\det(J)<0$. Diffeomorphisms that are connected to the identity, denoted by $\mathrm{Diff}_0(\mathcal{M})$, form a subgroup and the action of infinitesimal diffeomorphisms on fields is represented by the Lie derivative $\mathcal{L}_\xi$ with respect to the vector field $\xi^\mu$. If we assume that the spacetime manifold $\mathcal{M}$ is orientable and has a topology that is trivial \cite{nash2013topology}, we have then that the mapping class group of spacetime diffeomorphisms modulo transformations connected to the identity,
\begin{equation}
\mathrm{Diff}(\mathcal{M})/\mathrm{Diff}_0(\mathcal{M})\;=\;\{\Id,\mathcal{R}\}\,,
\label{mapping-class-group}
\end{equation}
is generated by the manifold orientation reversal $\mathcal{R}:\Xi\mapsto-\Xi$. As we have introduced no Lorentzian metric up to this point, there is no distinction between spatial parity and time reversal, only a notion of orientation reversal $\mathcal{R}$.

\medskip

In the Einstein-Cartan formulation of General Relativity, the fundamental dynamical variables are a Lorentz connection $\omega_\mu^{IJ}(x) dx^\mu$ and a coframe field $e^I_\mu(x)dx^\mu$. The Lorentz connection allows us to parallel transport Lorentz fields. The coframe field allows us to define a Lorentzian metric as a derived quantity.  The ordered basis $E_I=E_I^\mu(x)\partial_\mu$ discussed earlier, the frame field, is also a derived quantity defined by $ \langle e^I,E_J\rangle=\delta^I{}_J$, i.e., the change of basis $E_I^\mu$ from the coordinate frame $\partial_\mu$ to the non-coordinate frame $E_I$ is given by the inverse of the coframe field, $E_I^\mu=(e_\mu^I)^{-1}$. The requirement that the two basis represent the same orientation implies that the determinant of the coframe field is positive,
\begin{equation}
\Xi=[\partial_\mu]=[E_I]\quad\Longrightarrow\quad \det(e_\mu^I)>0\,.
\end{equation} 
Equivalently, we can consider the volume $4$-form 
\begin{equation}
\Omega_{\mathrm{v}}\,=\,\frac{1}{4!}\,\epsilon_{IJKL}\,e^I\wedge e^J\wedge e^K\wedge e^L\,,
\label{OmegaV}
\end{equation}
together with the coordinate $4$-form 
\begin{equation}
\dd^4x\,=\,\frac{1}{4!}\,\epsilon_{\mu\nu\rho\sigma}\,dx^\mu \wedge dx^\nu \wedge dx^\rho\wedge dx^\sigma\,,
\end{equation}
where $\epsilon_{IJKL}$ and $\epsilon_{\mu\nu\rho\sigma}$ are alternating symbols (with $\epsilon_{0123}=+1$). As they are both nowhere-vanishing top forms and $\Omega_{\mathrm{v}}=\det(e_\mu^I)\,\dd^4x$, we have again
\begin{equation}
\Xi=[\dd^4x]=[\Omega_{\mathrm{v}}]\quad\Longrightarrow\quad \det(e_\mu^I)>0\,.
\end{equation}
We turn now to the discussion of diffeomorphisms, and in particular of orientation-reversing diffeomorphisms, in the Einstein-Cartan formulation. Let us consider a Lorentz vector-valued one-form $\alpha^I=\alpha^I_\mu(x)dx^\mu$. Given a spacetime diffeomorphism $\varphi\in\mathrm{Diff}(\mathcal{M})$, we need to define a Kosmann lift to the principle Lorentz bundle \cite{Kosmann,Fatibene2011,Jacobson:2015uqa,Prabhu:2015vua},
\begin{equation}
\hat{\varphi}\;\in \; O(1,3)_{\mathcal{M}}\rtimes\mathrm{Diff}(\mathcal{M})\,,
\label{Lorentz-bundle}
\end{equation}
that allows us to compare Lorentz vectors at different spacetime points, i.e.,
\begin{equation}
\alpha^I_\mu(x)\quad\longrightarrow\quad (\Lambda_\varphi(x))^I{}_K\;\alpha^K_\nu(\varphi(x))\;(J_\varphi(x))^\nu{}_\mu\,.
\end{equation}
At the infinitesimal level, the lift of a spacetime diffeomorphism to Lorentz vectors is given by the Lorentz-covariant Lie derivative $\hat{\mathcal{L}}_\xi$. This notion requires the use of the Lorentz connection $\omega_\mu^{IJ}$, with
\begin{equation}
\hat{\mathcal{L}}_\xi \alpha^I_\mu\;=\;\xi^\nu\partial^{\vphantom{I}}_\nu \alpha^I_\mu + \alpha^I_\nu\partial^{\vphantom{I}}_\mu \xi^\nu  \;+\;\xi^\nu\omega^{\,I}_{\nu\, J}\,\alpha^J_\mu\,.
\end{equation}
For large diffeomorphims that are not connected to the identity, the natural lift $\hat{\varphi}$ extends consistently to reversals $\mathcal{R}$ of the orientation defined in a coordinate basis $\Xi=[\partial_\mu]=[\dd^4x]$ or equivalently in a non-coordinate basis $\Xi=[E_I]=[\Omega_{\mathrm{v}}]$, i.e.,
\begin{align}
\hat{\mathcal{R}}: \;\;&[\dd^4x]\,\mapsto -[\dd^4x]\,,\\[.5em]
&[\Omega_{\mathrm{v}}]\;\:\mapsto -[\Omega_{\mathrm{v}}]\,.
\end{align}
This natural lift of the spacetime orientation reversal requires that the local Lorentz transformation $\Lambda_\varphi$ has the same determinant sign as the Jacobian $(J_\varphi)^\mu{}_\nu=\partial \varphi^\mu/\partial x^\nu$,
\begin{equation}
\det( \Lambda_\varphi)\;=\;\sgn \det(J_\varphi)\,.
\end{equation}
Specifically, an orientation reversing diffeomorphism $\det(J_\varphi)<0$ always comes together with an internal space Lorentz reversal, $\det( \Lambda_\varphi)=-1$. In particular, we have that under a spacetime orientation-reversing diffeomorphism $\hat{\varphi}$, the coframe field transforms as $e_\mu^I\to (\Lambda_\varphi)^I{}_K\;e^K_\nu\;(J_\varphi)^\nu{}_\mu$ and, as a result, $\det(e)>0$ is sent to $\det(\Lambda_\varphi\, e\, J_\varphi)>0$\,. With these preliminaries, we are ready to discuss how this lift provides a consistent definition of spacetime orientation reversals both in the first-order and in the second-order formulations of General Relativity.

\medskip

Einstein's general covariance is realized in General Relativity as the invariance of the action $S=\int\Omega$ under spacetime diffeomorphisms. While diffeomorphisms connected to the identity are assumed to be an exact symmetry also at the quantum level, transformation in the mapping class group \eqref{mapping-class-group} are expected to be only accidental approximate symmetries in the semiclassical regime. This is what happens for instance in $2+1$ quantum gravity \cite{Carlip:1998uc} and in LQG in four dimensions \cite{RovelliW-EDiscrete}. As we assume here that the effective action is invariant only under $\mathrm{Diff}_0(\mathcal{M})$, it is useful to list the transformation properties of various $4$-forms $\Omega$ that can appear in the action for gravity. We start from the coframe field $e^I$,  the coframe torsion $\tau^I$, and the Lorentz curvature $F^{IJ}$,
\begin{align}
\tau^I= &\;\, \nabla e^I\,=\,d e^I+\omega^I{}_J\,e^J\,,\\
F^{I}{}_{J}= &\;\, d\omega^{I}{}_{J}+\omega^I{}_K\wedge \omega^{K}{}_{J}\,,
\end{align}
and we raise and lower Lorentz indices $I, J,\ldots$ with the internal Minkowski metric $\eta_{IJ}$. The only $4$-form that is Lorentz invariant, polynomial in $e^I$ and $\omega^{IJ}$, and contains no exterior derivatives is the volume $4$-form $\Omega_{\mathrm{v}}$ \eqref{OmegaV}. With one single exterior derivative we have the Einstein-Cartan and the Holst $4$-forms
\begin{align}
\Omega_{EC}\;=\;&\;\frac{1}{2}\epsilon_{IJKL}\,e^I\wedge e^J\wedge F^{KL}\,,\label{OmegaEC}\\[.5em]
\Omega_{H}\;\;= \;&\;e_I\wedge e_J\wedge F^{IJ}\,,\label{OmegaH}
\end{align}
With two exterior derivatives, we have the torsion-squared $4$-form $\Omega_{TT}$, the Gauss-Bonnet (also known as Euler) $4$-form $\Omega_{GB}$ and the Chern-Simons (also known as Pontryagin) $4$-form $\Omega_{CS}$ \cite{RezendePerezTopoterms,Baekler:2011jt,NYRomesh,Mercuri:2007ki, NYBanerjee}:
\begin{align}
\Omega_{TT}\;=\;&\;\nabla e_I\wedge \nabla e^I\,,\label{OmegaTT}\\[.5em]
\Omega_{GB}\;=\;&\;\frac{1}{2}\epsilon_{IJKL}\,F^{IJ}\wedge F^{KL}\,,\label{OmegaGB}\\[.5em]
\Omega_{CS}\;=\;&\;F_{IJ}\wedge F^{IJ}\,\label{OmegaPCS}.
\end{align}
Under orientation-reversing diffeomorphisms
$\mathcal{R}$, we have the transformation properties listed in Table~\ref{table:R}.


\begin{table}[t]
    \centering
\begin{tabular}{ |c|c|c| } 
\hline
 & Odd & Even \\
\hline
\,\,$\mathcal{O}(F^0)$\,\, & $\,\Omega_{\mathrm{v}}\,\overset{\mathcal{R}}{\longrightarrow}\, -\Omega_{\mathrm{v}}\,$ & $\,\Omega_{TT}\,\overset{\mathcal{R}}{\longrightarrow}\, +\Omega_{TT}\,$\\
\hline
$\mathcal{O}(F^1)$ & $\,\Omega_{EC}\,\overset{\mathcal{R}}{\longrightarrow}\, -\Omega_{EC}\,$ & $\,\Omega_{H}\,\overset{\mathcal{R}}{\longrightarrow}\, +\Omega_{H}\,$\\
\hline
$\mathcal{O}(F^2)$ & $\,\Omega_{GB}\,\overset{\mathcal{R}}{\longrightarrow}\, -\Omega_{GB}\,$ & $\,\Omega_{CS}\,\overset{\mathcal{R}}{\longrightarrow}\, +\Omega_{CS}\,$ \\
\hline
\end{tabular}
\caption{\comment{Transformation properties of the gravitational action forms under orientation reversing diffeomorphisms $\mathcal{R}$.}}
    \label{table:R}
\end{table}

Therefore, the Einstein-Cartan action with a cosmological constant, $S_{EC\Lambda}=\frac{1}{2\kappa}\int (\Omega_{EC}-2\Lambda\, \Omega_{\mathrm{v}})$ is odd under orientation reversals, $S_{EC\Lambda}\to -S_{EC\Lambda}$. On the other hand, including also a Holst term or a Chern-Simons term results in an action that is neither odd nor even under orientation reversals. We will call an action that is neither odd nor even under orientation-reversing diffeomorphisms a \emph{parity-violating action} for short.

\bigskip

The conditions for a parity-violating action described above apply also to the metric formulation of General Relativity. Let us assume that the spacetime torsion vanishes, $\tau^I=\nabla e^I=0$, and that there is no coupling to spinor fields that can source torsion. We can then formulate the theory purely in terms the spacetime metric $g_{\mu\nu}(x)$ taken as a fundamental field, instead of considering it a derived quantity
\begin{equation}
g_{\mu\nu}(x)\,=\,\eta_{IJ}^{\vphantom{I}}\,e^I_\mu(x)e^J_\nu(x)\,,
\end{equation}
as done in the Einstein-Cartan formulation. Moreover the condition of zero torsion $\tau^I=0$ allows us to solve for the metric-compatible Lorentz connection that defines the Christoffel connection $\Gamma^{\rho}_{\mu\nu}=\Gamma^{\rho}_{\mu\nu}(g)$ and the Riemann tensor $R^{\mu}{}_{\nu\rho\sigma}(g)$. In the metric formulation, the volume $4$-form \eqref{OmegaV} can be expressed as
\begin{equation}
\Omega_{\mathrm{v}}=\sqrt{-g}\,\dd^4x
\end{equation}
in terms of the coordinate $4$-form $\dd^4x=dx^0\wedge dx^1\wedge dx^2\wedge dx^3$ and the volume density $\sqrt{-g}=\sqrt{-\det g_{\mu\nu}}$. Note that we have already assumed that the coordinate basis and the coframe field define the same orientation, and therefore 
$\sqrt{-g}=\det(e_\mu^I)>0$. We can now write the $4$-forms described earlier in terms of the metric, under the assumption of vanishing torsion. In particular we have that the Einstein-Cartan $4$-form reduces to the Ricci scalar times the volume $4$-form (also known as the Einstein-Hilbert density), while the Holst $4$-form vanishes:
\begin{align}
\Omega_{EC}^{(\tau=0)}\;=\;&\;R\, \sqrt{-g}\,\dd^4x\,,\label{Omega-EC-metric}\\[.5em]
\Omega_{H}^{(\tau=0)}\;=\;&\;0\,.
\end{align}
The second equation follows from the identity $\Omega_{H}=\Omega_{TT}-\Omega_{NY}$ where $\Omega_{NY}=d(e_I\wedge \tau^I)$ is the Nieh-Yan $4$-form. Similarly, the Gauss-Bonnet and the Chern-Simons $4$-forms in the absence of torsion reduce to
\begin{align}
\Omega_{GB}^{(\tau=0)}=&\;\tfrac{1}{2}\big(R_{\mu\nu\rho\sigma}R^{\mu\nu\rho\sigma}-4R_{\mu\nu}R^{\mu\nu}+R^2\big)\, \sqrt{-g}\,\dd^4x\,,\nonumber\\[.5em]
\Omega_{CS}^{(\tau=0)}=&\;\tfrac{1}{4}\,\epsilon^{\mu\nu\rho\sigma}\,R^\alpha{}_{\beta\mu\nu}\,R^\beta{}_{\alpha\rho\sigma}\;\dd^4x\,. \label{Omega-CS-metric}
\end{align}
In the metric formulation, the Lagrangian $L$ associated to a $4$-form $\Omega$ is defined as
\begin{equation}
\Omega\,=\,L \,\sqrt{-g}\,\dd^4x\,.
\label{Omega-Lagrangian}
\end{equation}
Orientation-reversing diffeomorphisms in the metric formulation do not require a lift from spacetime to the Lorentz bundle \eqref{Lorentz-bundle}. The transformation properties of the metric $4$-forms (\ref{Omega-EC-metric}-\ref{Omega-CS-metric}) coincide with (Table \ref{table:R}). Note that, as $\dd^4x\to -\dd^4x$ under orientation reversals, we have that the action $S=\int \Omega$ and the associated Lagrangian $L$ have opposite transformation properties
\begin{equation}
S\,\overset{\mathcal{R}}{\longrightarrow}\, \mp S\quad\Longrightarrow\quad L\,\overset{\mathcal{R}}{\longrightarrow}\,\pm L\,,
\end{equation}
i.e., the Lagrangian $L$ transforms as a scalar or a pseudoscalar. In particular, the Einstein-Hilbert action $S_{EH}=\frac{1}{2\kappa}\int R\, \sqrt{-g}\,\dd^4x$ is odd under orientation reversals, while the Einstein-Hilbert Lagrangian $L_{EH}=\frac{1}{2\kappa} R$ is even and therefore a scalar. 

\medskip

In this analysis, we have focused on orientation-reversing spacetime diffeomorphisms and their lift to the principle Lorentz bundle. We have not discussed purely internal Lorentz transformations that are Lorentz reversals, such as the internal parity $P$ and internal time reversal $T$ considered in \cite{RovelliW-EDiscrete}. These transformations change the determinant of the frame field $e_\mu^I$ without changing the orientation of the spacetime manifold, and therefore they do not have a corresponding transformation in the metric formulation.\\

In formal manipulations in metric variables, it is useful to adopt the metric-dependent antisymmetric tensor \cite{Carrollbook},
\begin{equation}
\varepsilon_{(g)}^{\mu\nu\rho\sigma}=\frac{-1}{\sqrt{-g}}\,\epsilon^{\mu\nu\rho\sigma}\,,
\label{epsilon-g}
\end{equation}
which transforms as a tensor and is defined in terms of the Levi-Civita alternating symbol $\epsilon^{\mu\nu\rho\sigma}$ (with $\epsilon^{0123}=+1$) appearing for instance in \eqref{Omega-CS-metric}.

\medskip

We include here a remark about the Gauss-Bonnet and Chern-Simons densities that will be relevant in the next section. The integral of $\Omega_{GB}$ and of $\Omega_{CS}$ defines topological invariants of a manifold as they are given by the exterior derivatives of $3$-forms,
\begin{align}
\Omega_{GB}=&\;d\big(\tfrac{1}{2}\epsilon_{IJKL}\,\omega^{IJ}\wedge \big(F^{KL}-\tfrac{1}{3}\omega^K{}_M\wedge\omega^{ML}\big)\big),\\[.5em]
\Omega_{CS}=&\;d\big(\omega_{IJ}\wedge \big(F^{IJ}-\tfrac{1}{3}\omega^I{}_M\wedge\omega^{MJ}\big)\big)\,.
\end{align}
In metric variables
\begin{eqnarray}
L_{GB}=\nabla_\mu K_{GB}^\mu \\
L_{CS}=\nabla_\mu K_{CS}^\mu
\end{eqnarray}
where $K_{GB}^\mu$ is given by \cite{CapozzielloCherubini}
\begin{equation}\label{GB current}
K_{GB}^\mu=\varepsilon_{(g)}^{\mu\nu\rho\sigma}\varepsilon_{(g)}^{\alpha\beta}{}_{\gamma\delta}\,\Gamma^\gamma_{\alpha\nu}\big(\tfrac{1}{2}R^\delta{}_{\beta\rho\sigma}+\tfrac{1}{3}\Gamma^\delta_{\lambda\rho}\Gamma^\lambda_{\beta\sigma} \big)\, ,
\end{equation}
and  $K_{CS}^\mu$ is the Chern-Simons current, which in the absence of torsion can be written as \cite{AlexanderYunes, JackwiPi}
\begin{equation}\label{P current}
K_{CS}^\mu=\varepsilon_{(g)}^{\mu\nu\rho\sigma}\,\Gamma^\alpha_{\nu\beta}\big(\partial_\rho\Gamma^\beta_{\sigma\alpha}+\tfrac{2}{3}\Gamma^\beta_{\rho\lambda}\Gamma^\lambda_{\sigma\alpha}\big),
\end{equation}
Even though by themselves they do not contribute to the equations of motion as they are total derivatives, a non-minimal coupling of the form \eqref{L_GB}, \eqref{L_CS} makes them no longer trivial.

\section{\boldmath{$\gamma$}-Dual Effective Action}\label{sec:Model}
In this section we describe the construction of an effective action for inflation motivated by the $\gamma$-duality of the spinfoam dynamics. The $\gamma$-dual action $S_\gamma$ is obtained via a duality rotation of a seed action $S_{\mathrm{odd}}$ that is parity-non-violating (\emph{odd} under orientation-reversing diffeomorphisms $\mathcal{R}$). While, the action $S_{\mathrm{odd}}$ we start with here is chosen with the requirement that it provides a minimal  model of inflation compatible with current observations \cite{PlanckX}, the construction of a $\gamma$-dual action via a duality rotation is more general and applies also to other possible seed actions.

\medskip

We consider an action for gravity and a single scalar field with a potential that drives inflation, written in first order formalism:
\begin{equation}
S_{\mathrm{odd}}^{(1)}=\int \tfrac{1}{2\kappa}\Omega_{EC}\,+\, L_m\, \Omega_{\mathrm{v}} \,+\,\Omega_{S}\,.
\end{equation}
The action depends on $e^I$ and $\omega^{IJ}$, the Lorentz coframe and the Lorentz connection; and on $\phi$, and $\pi^I$, the scalar field and its covariant momentum. It is given by the integral of the Einstein-Cartan $4$-form $\Omega_{EC}$ \eqref{OmegaEC},   the volume $4$-form $\Omega_{\mathrm{v}}$ \eqref{OmegaV},  and  the $4$-form 
\begin{equation}
\Omega_S=\frac{1}{3!}\epsilon_{IJKL}\,\pi^I\, d\phi\wedge e^J\wedge e^K\wedge e^L\,,\label{OmegaS}
\end{equation} 
that depends both on the scalar field $\phi$ and its covariant momentum, the vector valued scalar $\pi^I$. The dynamics of the scalar field is encoded in the Lagrangian $L_m$,
\begin{equation}
L_m=+\tfrac{1}{2}\pi^I \pi_I-V(\phi)\,,
\end{equation}
with a potential $V(\phi)$ with a plateau that can drive slow-roll inflation \cite{PlanckX}, such as the Starobinsky potential \cite{StarobinskyPotential}.
This first-order action is equivalent to the one generally considered in single-field slow-roll inflation in the metric formalism: the stationarity of the action under variations of the Lorentz connection $\omega^{IJ}$ and of the scalar momentum $\pi^I$ imposes that the torsion vanishes $\tau^I=\nabla e^I=0$ and the momentum is given by $\pi_I=e_I^\mu\partial_\mu \phi$. \comment{Working in the Einstein-Cartan formalism, though, allows us to highlight three properties of this action that are relevant for our construction:

(\emph{i}) The action $S_{\mathrm{odd}}^{(1)}$ is polynomial in the one-forms $e^I$ and $\omega^{IJ}$. In particular, it does not require that the coframe field $e^I_\mu(x) dx^\mu$ has an inverse $e^\mu_I=(e^I_\mu)^{-1}$. As such, it allows spinfoam-like $2$d configurations of the geometry.

(\emph{ii}) In an expansion in the number of exterior derivatives $d$ of fields, the term $L_m\, \Omega_{\mathrm{v}}$ is of zero-th order, while the two terms $\Omega_{EC}$ and $\Omega_S$ are of first order as they depend on $d\omega^{IJ}$ and $d\phi$.

(\emph{iii}) The action defines a theory that is invariant under spacetime diffeomorphisms $\mathrm{Diff}(\mathcal{M})$, including orientation-reversing diffeomorphisms $\mathcal{R}$ as the action is parity-odd, $S_{\mathrm{odd}}^{(1)} \overset{\mathcal{R}}{\longrightarrow} -S_{\mathrm{odd}}^{(1)}$. 

\medskip

\noindent This last property (iii) also holds in the metric formalism (the Einstein-Hilbert action with minimally coupled inflaton).
Following Weinberg's construction of an effective field theory of inflation \cite{Weinberg}, we consider now an extension $S_{\mathrm{odd}}^{(2)}$ of the action $S_{\mathrm{odd}}^{(1)}$ that includes terms of second order in exterior derivatives, 
with the additonal condition that it is (i) polynomial in the one-forms $e^I$ and $\omega^{IJ}$, and (iii) parity odd under orientation-reversing diffeomorphisms. These conditions exclude $\Omega_{TT}$ and $\Omega_{CS}$, the torsion-squared and the Chern-Simons $4$-forms which are parity-even (Table. \ref{table:R}).} 
We are left with the Gauss-Bonnet $4$-form and $S_{\mathrm{odd}}^{(2)}=\int f(\phi)\, \Omega_{GB}$,  where $f(\phi)$ is a function of the scalar field $\phi$. Note that we do not consider possible couplings to the covariant momentum $\pi^I$ which effectively correspond to a modified kinetic term for the scalar field as in models of $K$-inflation \cite{Armendariz-Picon:1999hyi}.


The parity non-violating action $S_{\mathrm{odd}}=S_{\mathrm{odd}}^{(1)}+S_{\mathrm{odd}}^{(2)}$,
\begin{equation}
S_{\mathrm{odd}}=\int \tfrac{1}{2\kappa}\Omega_{EC}\,+\, L_m\, \Omega_{\mathrm{v}} \,+\,\Omega_{S}\,+\,f(\phi)\, \Omega_{GB}\,,\label{S-odd}
\end{equation}
is taken here as the starting point, or seed, for the construction of a $\gamma$-dual action for inflation. Under a duality rotation \eqref{eq:duality-rotation-def} of the Lorentz curvature $F^{IJ}$, the $4$-forms appearing in $S_{\mathrm{odd}}$ transform as\footnote{These transformation properties can equivalently be derived using the self-dual and antiself-dual variables $F^{(\pm)}_{IJ}=\frac{1}{2}(F_{IJ}\mp\ii\, {}^*\!F_{IJ})$ which transform as $F^{(\pm)}_{IJ}\overset{\theta}{\longrightarrow} \ee^{\pm \ii\theta}F^{(\pm)}_{IJ}$.} 
\begin{align}
\Omega_{\mathrm{v}}\;\;\,\quad&\overset{\theta}{\longrightarrow}\quad \Omega_{\mathrm{v}}\,,\label{theta-V}\\
\Omega_{S}\;\;\quad&\overset{\theta}{\longrightarrow}\quad \Omega_{S}\,,\label{theta-S}\\
\Omega_{EC}\quad&\overset{\theta}{\longrightarrow}\quad \cos\theta \;\;\,\Omega_{EC} \;-\;\sin\theta\;\;\, \Omega_{H}\,,\label{theta-EC}\\
\Omega_{GB}\quad&\overset{\theta}{\longrightarrow}\quad \cos2\theta \;\Omega_{GB}\; -\;\sin2\theta\; \Omega_{CS}\,,\label{theta-GB}
\end{align}
and define a one-parameter family of duality-rotated actions $S_{\mathrm{odd}}\overset{\theta}{\longrightarrow} S_\theta$. 

\medskip

We note that duality rotations of the curvature tensor have been considered multiple times in the literature with the different goal of relating solutions of Einstein equations \cite{penrose_rindler,HenneauxTeitelboim}.\footnote{Note that here we consider only duality rotations in the Lorentz internal indices of $F^{IJ}$. Alternatively, in the metric formalism, one can also consider a duality rotation of the Riemann tensor on spacetime manifold indices (via a tangent-space Hodge dual operator). These two definitions are not trivially equivalent; for a discussion on this comparison, see \cite{Thesis}. }  Recently, the duality rotation \eqref{theta-EC} of the Einstein-Cartan action has been considered in \cite{Kol:2022bsd,Kol:2023yxd} as an exact symmetry of vacuum Einstein gravity, but shown in \cite{Monteiro:2023dev} to be only a symmetry of a duality-preserving sector of vacuum solutions. The transformation considered here is not a symmetry of the action $S_{\mathrm{odd}}$, but the construction of a one-parameter family of actions $S_\theta$ that reproduces the relation between the EPRL spinfoam model and the parity-non-violating limit $\theta\to 0$ given by the BC model discussed in Sec.~\ref{sec: gDuality Spinfoams}. By setting the parameter $\theta$ to the value $\tan\theta=1/\gamma$ that relates it to the Barbero-Immirzi parameter $\gamma$ \eqref{eq:theta-gamma} and normalizing the constant $\kappa/\cos\theta=8\pi G$ to the observed value of  Newton's constant $G$ as discussed in Sec.~\ref{sec: gDuality Spinfoams}, we obtain the spinfoam-motivated $\gamma$-dual action,
\begin{equation}
S_{\gamma}=S_{\gamma}^{(1)}+S_{\gamma}^{(2)}
\label{Sgamma}
\end{equation}
with 
\begin{widetext}
\begin{align}
S_{\gamma}^{(1)}[e^I,\omega^{IJ},\phi\,]
&=\frac{1}{16 \pi G}\!\int \left(\frac{1}{2}\epsilon_{IJKL}\,e^I\wedge e^J\wedge F^{KL}- \frac{1}{\gamma}\, e_I\wedge e_J\wedge F^{IJ}\right) \nonumber\\
&\quad +\int \left(-\frac{1}{2}\,\eta^{IJ}e^\mu_I\partial_\mu\phi\; e^\nu_J\partial_\nu\phi-V(\phi)\right)\frac{1}{4!}\,\epsilon_{IJKL}\,e^I\wedge e^J\wedge e^K\wedge e^L\,\label{Sgamma1} \\[.5em]
S_{\gamma}^{(2)}[e^I,\omega^{IJ},\phi\,]&=\int f(\phi)\left(\frac{\gamma^2-1}{\gamma^2+1}\;\frac{1}{2}\epsilon_{IJKL}\,F^{IJ}\wedge F^{KL}\,-\frac{2\gamma}{\gamma^2+1}F_{IJ}\wedge F^{IJ}\right) \,,\label{Sgamma2}
\end{align}
\end{widetext} 
where we have already solved for the covariant momentum $\pi^I$. Furthermore, imposing the zero-torsion condition $\tau^I=0$ as a constraint on the Lorentz connection,\footnote{Alternatively, one can implement a reduction of order procedure where one substitutes the solutions of the equations of motion of the first-order action $S_{\gamma}^{(1)}$ into the second order action $S_{\gamma}^{(2)}$ as discussed in \cite{Weinberg}.} we can write the $\gamma$-dual action in the metric formalism
\begin{widetext}
\begin{align}\label{Seff gDual}
S_{\gamma}[ g_{\mu\nu},\phi]
=&\;\; \frac{1}{16\pi G}\!\int R \,\sqrt{-g}\,\dd^4x \;\; +\int \Big(-\frac{1}{2}\,g^{\mu\nu}\partial_\mu\phi\, \partial_\nu\phi-V(\phi)\Big)\sqrt{-g}\,\dd^4 x \\
& +\int f(\phi)\left(\frac{\gamma^2-1}{2(\gamma^2+1)}\left(R_{\mu\nu\rho\sigma}R^{\mu\nu\rho\sigma}-4R_{\mu\nu}R^{\mu\nu}+R^2\right)\,+\frac{\gamma}{2(\gamma^2+1)}\varepsilon_{(g)}^{\alpha\beta\rho\sigma}R_{\mu\nu\alpha\beta}R^{\mu\nu}{}_{\rho\sigma}\right)\sqrt{-g}\,\dd^4 x \,.\nonumber
\end{align}
\end{widetext} 
In the remaining part of the paper we study primordial cosmological perturbations described by this parity-violating $\gamma$-dual action and the possibility of determining the Barbero-Immirzi parameter from observations.

Parity violating actions of the form \eqref{Seff gDual} have been extensively studied \cite{LueWangKamionkowski,AlexanderYunes,JackwiPi,Alexander:2004us,Alexander:2004wk,Lyth:2005jf,ContaldiMagueijoSmolin,SatohSoda,Satoh2010,Dyda:2012rj,Kawai:2017kqt,Gialamas:2022xtt,Bartolo:2014hwa,Bartolo:2015dga,Bartolo:2017szm,Nojiri:2019nar,Bordin:2020eui,Alexander:2022cow,Gong:2023kpe,Jenks:2023pmk,Creque-Sarbinowski:2023wmb}. Specifically, in \cite{SatohSoda,Satoh2010}  an action for inflation that includes both the Gauss-Bonnet and the Chern-Simons terms was considered, but with independent couplings $f_{GB}(\phi)$ and $f_{CS}(\phi)$ to the scalar field. The requirement of $\gamma$-duality motivated by the spinfoam dynamics fixes the ratio of these two functions to a constant related to the Barbero-Immirzi parameter, $f_{GB}(\phi)/f_{CS}(\phi)=\gamma-\frac{1}{\gamma}$.

We note that, up to two conditions, the effective action \eqref{Seff gDual}  coincides with the action describing an effective field theory of inflation discussed in \cite{Weinberg}. The logic there is to consider the most general action for a metric $g_{\mu\nu}$ and a scalar field $\phi$ that is invariant under diffeomorphisms, it is organized in the number of derivatives (truncated at the fourth derivative), up to field redefinitions and with reduction of the order with respect to the truncation to second derivatives. Using reduction of order with respect to the equations of motion of the action $S_\gamma^{(1)}$, allows one to write the Weyl-squared term $C_{\mu\nu\rho\sigma}C^{\mu\nu\rho\sigma}$ as a Gauss-Bonnet term, up to a redefinition of coupling functions $f(\phi)$. The first difference with respect to \cite{Weinberg} is that \eqref{Seff gDual} does not include a $K$-inflation type correction to the matter sector \cite{Armendariz-Picon:1999hyi},
 \begin{equation}\label{k-inflation correction}
 S_{K}=\int f_{K}(\phi)\left(g^{\mu\nu} \partial_\mu\phi\,\partial_\nu\phi\right)^2\,\sqrt{-g}\,\dd^4x\,.
 \end{equation}
In principle, this term can be included in our action but we do not explore it here. The second difference is that in \cite{Weinberg} there are two independent coupling functions $f_{GB}(\phi)$ and $f_{CS}(\phi)$. Here, the new condition of $\gamma$-duality relates the ratio of the two to the Barbero-Immirzi parameter.

\section{Inflation and Primordial Gravitational Waves}\label{sec: Inflation and PGW}
In this section we study the power spectrum of primordial gravitational waves in a model of inflation given by the $\gamma$-dual action \eqref{Seff gDual}. We refer to \cite{weinberg2008cosmology, baumann2022cosmology} for an introduction to cosmological perturbations and to \cite{LueWangKamionkowski,AlexanderYunes,JackwiPi,Alexander:2004us,Alexander:2004wk,Lyth:2005jf,ContaldiMagueijoSmolin,SatohSoda,Satoh2010,Dyda:2012rj,Kawai:2017kqt,Gialamas:2022xtt,Bartolo:2014hwa,Bartolo:2015dga,Bartolo:2017szm,Nojiri:2019nar,Bordin:2020eui,Alexander:2022cow,Gong:2023kpe,Jenks:2023pmk,Creque-Sarbinowski:2023wmb} for the study of primordial gravitational waves from parity-violating actions including a Chern-Simons term. 

\medskip

We consider a homogeneous and isotropic, spatially flat $(k=0)$ Friedman-Lemaître-Robertson-Walker (FLRW) background. Tensorial perturbations are described by the perturbed metric
\begin{equation}
g_{\mu\nu}dx^\mu dx^\nu=-dt^2+a(t)^2 \big(\delta_{ij}+2E_{ij}(t,\vec{x})\big)dx^idx^j\,,
\end{equation}
where $t$ is cosmic time, $a(t)$ the scale factor, the spatial indices $i,j,\ldots$ are raised and lowered with the $3d$ Euclidean metric $\delta_{ij}$, and $E_{ij}$ the transverse-traceless component in a scalar-vector-tensor decomposition of perturbations. In this section we do not consider scalar  perturbations, which decouple from tensor perturbations in the quadratic expansion of the action we are working in.

A 3-dimensional Fourier transform allows us to work directly with the two circular polarizations $(\pm)$ per mode,
\begin{equation}
E_{ij}(t,\vec{x})=\int\frac{\dd^3 \vec{k}}{(2\pi)^3}\,\ee^{\ii \vec{k}\cdot\vec{x}}\sum_{\lambda=\pm}J_{ij}^{\lambda}(\vec{k})\;h_\lambda \smalltk, 
\end{equation} 
where $J_{ij}^{\pm}(\vec{k})$ is the circular polarization basis with $k^i J_{ij}^{\pm}(\vec{k})=0$, $J_{ij}^{+}(-\vec{k})=J_{ij}^{-}(+\vec{k})$,  and normalization $\mathrm{Tr}( J^{+}(\vec{k})J^{-}(\vec{k}))=4$. 

\medskip
 
The equations of motion satisfied by the homogeneous and isotropic background solution $a(t)$, $\phi(t)$ are
\begin{align}
&H^2=\frac{\kappa}{3}\left(\frac{\dot\phi^2}{2}\,+\,V(\phi)\,+\,12\gbc\, H^3\,\dot{f}(\phi)\right)\\
&\ddot\phi+3H\dot\phi+V'(\phi)=12\gbc\left(H^4+H^2\dot H\right)f'(\phi),
\end{align}
where $H(t)=\frac{\dot a(t)}{a(t)}$ is the Hubble rate, a dot $(\,\dot{}\,)$ denotes cosmic time derivatives, and a prime $(\,'\,)$ denotes derivative with respect to the scalar field $\phi$.

To describe a slow-roll phase, one introduces the parameters
\begin{align}
&\epsilon_V(t)\equiv\frac{1}{2\kappa}\left(\frac{V'(\phi)}{V(\phi)}\right)^2 \;,\quad
\eta_V(t)\equiv\frac{1}{\kappa}\frac{V''(\phi)}{V(\phi)}\,,\\
&\epsilon_H(t)\equiv-\frac{\dot H}{H^2}\,,\quad
\delta(t)\equiv\frac{\ddot\phi}{\dot\phi H}\,,\quad
\rho^2(t)\equiv\frac{\kappa \dot\phi^2}{H^2}\,,
\end{align}
and demands that they are  small. 
The introduction of a coupling $f(\phi)$ and higher-order corrections to the action requires us to define additional slow-roll parameters. Following \cite{SatohSoda, Satoh2010}, we define
\begin{align}
\sigma(t)&\equiv\kappa \dot{f}(\phi) H(t)\label{sigma_def}\\
\alpha(t)&\equiv\frac{\kappa}{3}V'(\phi)f'(\phi)\\
\beta(t)&\equiv\frac{\kappa}{3} V(\phi)f''(\phi)\label{beta_def}
\end{align}
and assume that they are small to ensure that $f(\phi)$ varies slowly. Notice that some parameters, e.g., $\sigma,\,\alpha,\,\eta_V$, can be negative. 
  
All parameters are not just required to be small, but also to vary slowly in time. This implies a tower of slow-roll parameters, for each one of the quantities described above, defined as  
\begin{equation}
\sigma_2(t)\equiv-\frac{\dot\sigma(t)}{H(t)\sigma(t)}\;, \;\;\sigma_3(t)\equiv-\frac{\dot\sigma_2(t)}{H(t)\sigma_2(t)}\;,\,\ldots
\end{equation}
and similarly for $\{\epsilon_H,\epsilon_{H2},\epsilon_{H3}, \ldots\}$, etc., such that, all of them are small. We will collectively track the order of the slow-roll expansion with their maximum, $\epsilon\equiv \mathrm{max}( \epsilon_H(t),...,|\eta_V(t)|,...,|\sigma(t)|,|\sigma_2(t)|,...)\ll 1$.

Note that the slow-roll parameters are not independent from each other, for example,
\begin{align}
\epsilon_H(t)&=\epsilon_V(t)-2\gbc\alpha(t) \\
\delta(t)&=\epsilon_H(t)-\eta(t)+4\gbc(\beta+2\alpha)\\
\sigma_2(t)&=\epsilon_H(t)\left(1+2\frac{\beta(t)}{\alpha(t)}\right)+2\beta(t)\gbc-\delta(t)
\end{align}
and of particular importance to us,
\begin{equation}\label{sigma of alpha}
\sigma(t)=-\alpha(t)+2\gbc\frac{\alpha^2(t)}{\epsilon_V(t)}.
\end{equation}
This relationship follows from the fact that the functional form of our coupling for the Gauss-Bonnet and the Chern-Simons terms has the same function $f(\phi)$. For other consistency conditions, see \cite{SatohSoda}.

We can then write the background equations at the lowest order as
\begin{align}
H^2(t)&=\frac{\kappa}{3}\,V(\phi)+\mathcal{O}(\epsilon)\,,\\
\dot\phi(t)&=-\frac{V'(\phi)}{3 H(t)}\left(1-\frac{4\alpha}{\epsilon_V}\gbc\right)+\mathcal{O}(\epsilon).\label{phi-eq SR}
\end{align}
Here we see how there is an effective modification from $f(\phi)$ to the potential $V(\phi)$, with the appearance of the extra (constant to lowest order) $\alpha$-term, coming from the Gauss-Bonnet modification to the action. In a specific inflationary model for $V(\phi)$, equation (\ref{phi-eq SR}) can provide a constraint for $\sigma \gbc$: For example, a model with a coupling function such that $f'(\phi)>0$, and $V'(\phi)>0$, such as the Starobinsky potential \cite{StarobinskyPotential},
\begin{equation}
V_S(\phi)=\Lambda_S^4\left(1-\ee^{-\sqrt{\frac{2\kappa}{3}}\phi}\right)^2
\end{equation}
would require
\begin{eqnarray}
|\sigma|\,\gbc<\frac{\epsilon_H}{4},
\end{eqnarray} 
in order for the field to consistently roll down the potential ($\dot\phi<0$).

\medskip

The next step is to expand the action \eqref{Seff gDual} around the slow-roll background, up to quadratic order in tensorial perturbations $h_{\pm}(t,\vec{k})$. We find
\begin{equation}\label{Seff hpm}
S_{\gamma\pm}[h_\pm]=\frac{1}{2}\int\! dt\!\!\int\!\frac{\dd^3\vec{k}}{(2\pi)^3}\,a^3\Big(Z_\pm\,|\dot{h}_\pm|^2-\frac{k^2}{a^2}B_\pm|h_\pm|^2\Big),
\end{equation}
with
\begin{align}
&Z_\pm(t,k)=\textstyle\frac{4}{\kappa}\Big(1+4\kappa\dot{f}(\phi)\Big(\gbc H(t) \pm \pc \frac{k}{a(t)}\Big)\Big)\label{Apm}\\
&B_\pm(t,k)=\textstyle\frac{4}{\kappa}\Big(1+4\kappa\Big(\gbc\ddot{f}(\phi)\pm\pc\dot{f}(\phi)\frac{k}{a(t)}\Big)\Big)\label{Bpm}
\end{align}
and $k=|\vec{k}|$. Note that the circular polarizations $\pm$ have different quadratic actions and dynamics. A constant function $f(\phi)$ turns the correction null and recovers the action for tensorial perturbations in standard general relativity where $Z_\pm=B_{\pm}=4/\kappa$. 

An expansion in the slow-roll parameters up to $\mathcal{O}(\epsilon)$ simplifies the functions (\ref{Apm}) and (\ref{Bpm}) as second derivatives of the coupling function are of order $\mathcal{O}(\epsilon^2)$,
\begin{align}
Z_\pm(t,k)=&\textstyle \frac{4}{\kappa}\left(1+4\sigma(t)\!\left(\gbc\pm\pc\frac{k}{a(t)H(t)}\right)\right)\,,\label{Zslow}\\
B_\pm(t,k)=&\textstyle \frac{4}{\kappa}\left(1\pm 4\sigma(t)\,\pc\frac{k}{a(t)H(t)}\right).\label{Bslow}
\end{align}
We note that, as in Chern-Simons gravity  \cite{AlexanderYunes,JackwiPi}, here there is a perturbative instability because the functions $Z_\pm$ and $B_\pm$ become negative at high momentum $k$  for one of the two circular polarizations \cite{Dyda:2012rj}. However the $k$-dependent terms in (\ref{Zslow},\ref{Bslow}) appear together with the slow-roll parameter $\sigma$ \eqref{sigma_def}, making the band of stable modes sufficiently large. In contrast, in the general case where there are independent coupling functions $f_{GB}(\phi)$ and $f_{CS}(\phi)$ in \eqref{Seff gDual}, the background dynamics does not depend on the Chern-Simons coupling and the range of stable modes $k$ depends on a separate condition to be imposed on the function $f_{CS}(\phi)$  \cite{SatohSoda, Satoh2010}.

We assume here that the value of $\gamma$ is finite and not as small as the  order of magnitude of the slow-roll parameters, or smaller, otherwise this expansion is not complete and we would have to consider corrections to the action of order $\epsilon^2$ to account for terms like $\sigma\pc\frac{k}{a(t)H(t)}$. We will assume from now on that $\gamma$ is not too small and the slow-roll approximation is valid at $\mathcal{O}(\epsilon)$.

The field operator $\hat{h}_\pm$ for the tensor perturbations of each circular polarization satisfies the equal-time commutation relations
\begin{equation}
\Big[\hat{h}_\lambda(t,\vec{k}),\frac{d}{dt}\hat{h}_{\lambda'}(t,-\vec{k}')\Big]=\frac{\ii\, \hbar}{a^3 Z_\lambda}\delta_{\lambda\lambda'}(2\pi)^3\delta^{3}(\vec{k}-\vec{k}')
\label{eq:h CCR}
\end{equation}
as follows from the canonical quantization of a theory with action \eqref{Seff hpm} that includes a factor $a^3 Z_\lambda$ in the kinetic term. A Fock representation of the field operators is given by the mode expansion
\begin{equation}
\hat{h}_\pm(t,\vec{k})=u_{\pm}(t,k)\,\hat{b}_\pm(\vec{k})+\bar{u}_{\pm}(t,k)\,\hat{b}^\dagger_\pm(-\vec{k})\,,
\end{equation}
where $\hat{b}_\pm$ are bosonic  annihilation operators with $[\hat{b}_\lambda(\vec{k}),\hat{b}_{\lambda'}(\vec{k}')]=0$ and
\begin{align}
\Big[\hat{b}_{\lambda\vphantom{\lambda'}}(\vec{k}),\hat{b}^\dagger_{\lambda'}(\vec{k}')\Big]=\delta_{\lambda\lambda'}(2\pi)^3\delta^3(\vec{k}-\vec{k}')\,,
\end{align}
while $u_\pm(t,k)$ are mode functions (with $\bar{u}$ denoting the complex conjugate). We assume the mode functions to depend only on $k=|\vec{k}|$ to describe a Fock vacuum $\hat{b}_{\pm}(\vec{k})|0 \rangle=0$ that has homogeneous and isotropic correlation functions. The mode functions $u_\pm(t,k)$ satisfy the canonical Wronskian conditions
\begin{equation}\label{wronskian u}
{u}_\pm(t,k) \dot{\bar{u}}_\pm(t,k) - \dot{u}_\pm(t,k) \bar{u}_\pm(t,k)  =\frac{\ii\,\hbar}{a(t)^3Z_\pm(t,k)},
\end{equation}
 that follows from the commutation relations (\ref{eq:h CCR}), and the equations of motion
 \begin{widetext}
 \begin{align}
&\ddot{u}_\pm(t,k)+\left(3H(t)+\frac{\dot{Z}_\pm(t,k)}{Z_\pm(t,k)}\right)\dot{u}_\pm(t,k)
+\frac{k^2}{a(t)^2}\frac{B_\pm(t,k)}{Z_\pm(t,k)}u_\pm(t,k)=0\,, \label{eom u ZB}\\[.5em]
&\ddot{u}_\smallpm(t,k)+3H(t)\left(1\mp\frac{4}{3}\sigma(t)\pc\frac{k}{a(t)H(t)}\right)\dot{u}_\smallpm(t,k)
+\frac{k^2}{a(t)^2}\left(1-4\sigma(t)\gbc\right) u_\smallpm(t,k)=0\,,\label{eom u SR}
\end{align}
\end{widetext}
 which we have written both in terms of the functions $Z_\pm$, $B_\pm$ and in a slow-roll expansion. A solution $u_\pm(t,k)$ of the equations (\ref{wronskian u}, \ref{eom u ZB}) that is ultraviolet adiabatic defines a Fock vacuum $|0 \rangle$. Here we are interested in an adiabatic vacuum that generalizes the Bunch-Davies condition for modes in a band $[k_{\min},k_{\max}]$ that includes the pivot mode $k_\ast$ at which to compute the amplitude and the tilt of the tensor power spectrum, with e.g. $k_\ast=0.05\,\mathrm{Mpc}^{-1}$ for Planck satellite's CMB measurements \cite{PlanckX}. It is useful then to change the time parametrization to $t\rightarrow \s= \frac{k_\ast}{a(t)H(t)} $. Denoting derivatives with respect to $\s$ with a prime, the equation of motion (\ref{eom u SR}) becomes,
\begin{widetext}
\begin{equation}
u_{k,\smallpm}''(\s)+\left(\frac{-2-2\epsilon_H(s)}{\s}\pm\frac{8\gamma}{\gamma^2+1}\sigma(s)\frac{k}{k_\ast}\right)u_{k,\smallpm}'(\s)+\frac{k^2}{k_\ast^2}\left(1+2\epsilon_H(s)-4\gbc\sigma(\s)\right)u_{k,\smallpm}(\s)=0\,.
\end{equation}
\end{widetext}
We then change variables to
\begin{align}
&u_{k,\smallpm}(\s)\rightarrow y_{k,\smallpm}(\s)=\sqrt{M_{k,\smallpm}(\s)}u_{k,\smallpm}(\s),\\
&\mathrm{with}\quad M_{k,\smallpm}(\s)=\frac{k_\ast^3(1-\epsilon_H(\s))Z_{k,\smallpm}(\s)}{\hbar H(\s)^2s^2},
\end{align}
after which, the equations of motion become
\begin{equation}\label{eom y Q}
y_{k,\smallpm}''(s)+Q_{k,\smallpm}(s)\, y_{k,\smallpm}(s)=0.
\end{equation}
with 
\begin{flalign}\label{Q general}
Q_{k,\smallpm}(\s)&=\frac{-2-3\epsilon_H(\s)}{\s^2}\pm\frac{8\gamma}{\gamma^2+1}\frac{k}{k_\ast}\frac{\sigma(\s)}{\s}\nonumber\\
&\qquad+\frac{k^2}{k_\ast^2}\left(1+2\epsilon_H(s)-4\gbc\sigma(s)\right),
\end{flalign}
and the canonical Wronskian condition becomes,
\begin{equation}\label{wronskian y}
y_{k,\smallpm}(s) \bar{y}'_{k,\smallpm}(s)-y'_{k,\smallpm}(s)\bar{y}_{k,\smallpm}(s)=\ii.
\end{equation}
As a check, we note that the equation of motion (\ref{eom y Q}) reduces to the standard equation for tensor modes in general relativity when $\sigma\rightarrow 0$, i.e.,

\comment{
\begin{equation}
y_{k}''(s)+\left(\frac{-2-3\epsilon_H(s)}{s^2}+(1+2\epsilon_H(s))\frac{k^2}{k_\ast^2}\right) y_{k}(s)=0.
\end{equation}
}
Since we are interested in a small band of modes around $k_\ast$, and with modes freezing at $s=1$, the slowly-varying Hubble rate and slow-roll parameters can themselves be expressed as a series on $s$. Following \cite{logxExpansion}, we introduce a $\log(s)$ expansion for  the Hubble rate and the slow-roll parameters, 
\begin{widetext}
\begin{align}
&H(s)=H_\ast\,\left(\vphantom{\frac{|}{|}}1+\Big(\epsilon_{H\ast}+\epsilon_{H\ast}^2+\mathcal{O}(\epsilon^3)\Big)\log(s)\right.\nonumber\\
& \left. \phantom{something long} +\frac{1}{2}\Big(\epsilon_{H\ast}(\epsilon_{H\ast}+\epsilon_{H2\ast})+\epsilon_{H\ast}^2(2\epsilon_{H\ast}+3\epsilon_{H2\ast})\mathcal{O}(\epsilon^4)\Big)\log^2(s)+...\right)\,,\label{logsH}\\
&\epsilon_H(s)=\epsilon_{H\ast}+\left(\epsilon_{H\ast}\epsilon_{H2\ast}+\mathcal{O}(\epsilon^3)\right)\log(s)+\Big(\frac{1}{2}\epsilon_{H\ast}\epsilon_{H2\ast}(\epsilon_{H2\ast}+\epsilon_{H3\ast})+\mathcal{O}(\epsilon^4)\Big)\log^2(s)+...
\end{align}
\end{widetext}
(where the asterisk denotes evaluation at $s=1$ and the tracking parameter $\epsilon$ is evaluated at freezing as well). For any slow-roll parameter, the coefficient for $\log^n(s)$ is at least of $\mathcal{O}(\epsilon^{n+1})$; and for the Hubble rate, of $\mathcal{O}(\epsilon^{n})$. We obtain then the equation of motion at $\mathcal{O}(\epsilon)$ around the freezing time of the mode $k_\ast$ to be of the form (\ref{eom y Q}) and
\begin{widetext}  
\begin{equation}
Q_{k,\smallpm}(\s)=\frac{-2-3 \epsilon_{H\ast}}{\s^2}\pm\frac{8\gamma}{\gamma^2+1}\frac{k}{k_\ast}\frac{\sigma_\ast}{\s}+\frac{k^2}{k_\ast^2}\left(1+2\epsilon_{H\ast}-4\gbc\sigma_\ast\right) \;\;+\,\mathcal{O}(\epsilon^2).
\end{equation}
Similarly, the function $M_{k,\pm}(s)$ takes the form

\begin{equation}\label{Mser}
M_{k,\pm}(s)=\frac{4k_\ast^3}{\hbar\kappa\, H_\ast^2 s^2}\left(1-\epsilon_{H\ast}\left(1+2\log(s)\right)+4\sigma_\ast\left(\gbc \pm \pc\frac{k}{k_\ast} s\right)\right)\;\;+\,\mathcal{O}(\epsilon^2).
\end{equation}

\end{widetext}

\noindent This equation of motion has the Whittaker functions as a basis of solutions, $\lbrace y_{1,\smallpm}(s),y_{2,\smallpm}(s)\rbrace $,
\begin{flalign}\label{Whittaker solutions}
y_{1,\smallpm}(s)&=\alpha\,\mathcal{M}_{\mp\ii\mu_0,\nu_0}\left(\ii\kappa_0 s\right)\,+\,\beta\,\mathcal{W}_{\mp\ii\mu_0,\nu_0}\left(\ii\kappa_0 s\right)&\\
y_{2,\smallpm}(s)&=\bar{y}_{1,\smallpm}(s),&
\end{flalign}
with
\begin{align}
\mu_0
&=\frac{4\,\gamma}{\gamma^2+1}\sigma_\ast+\mathcal{O}(\epsilon^2)\,,\\
\nu_0&
=\frac{3}{2}+\epsilon_{H\ast}+\mathcal{O}(\epsilon^2)\,,\\
\kappa_0
&=2 \,\frac{k}{k_\ast}\left(1+\epsilon_{H\ast}-2\gbc\sigma_\ast\right)+\mathcal{O}(\epsilon^2)\,.
\end{align}
The coefficients $\alpha$ and $\beta$ are fully determined by the canonical Wronskian condition (\ref{wronskian y}) and a choice of initial adiabatic Bunch-Davies-like vacuum state; i.e., such that its two-point vacuum correlation function matches the one of Minkowski on the infinite past,
\begin{flalign}
&\lim_{s\rightarrow \infty} \bar{y}_\pm(s)y_\pm(s)=\frac{{k_\ast}}{2k}\,,\\
&\lim_{s\rightarrow \infty} y_\pm(s)=\sqrt{\frac{{k_\ast}}{2k}}\ee^{\ii \frac{k}{k_\ast}s}.
\end{flalign}
These conditions determine the adiabatic solutions given by
\begin{flalign}\label{yad}
\yad(s)=\frac{\ee^{\mp\frac{\pi}{2}\mu_0}}{\sqrt{\kappa_0}}\,\mathcal{W}_{\mp\ii\mu_0,\nu_0}\left(\ii\kappa_0 s\right)\,.
\end{flalign}
Using the asymptotic behavior of the Whittaker $\mathcal{W}$ function, and defining $\yasy(s)\sim \yad(s)$ in the $s\to 0$ limit, we have
\begin{equation}
\yasy(s)=\ee^{\frac{\pi}{4}(\ii\mp 2\mu_0)}s^{-\nu_0}\left(\frac{(\ii\kappa_0)^{-\nu_0}\Gamma(2\nu_0)}{\Gamma(\text{\small$\frac{1}{2}$}+\nu_0\pm\ii\mu_0)}s^{\frac{1}{2}}+\mathcal{O}(s^{\frac{3}{2}})\right),
\end{equation}
where $\Gamma$ is the Gamma function. With this expression and the  function (\ref{Mser}), we can compute the power spectrum $\Delta^2_\pm(k)$ of circularly polarized tensor perturbations,
\begin{align}\label{Power Dimless}
\Delta^2_\pm(k)&\equiv \lim_{t\to\infty}\frac{k^3}{2\pi^2}\left|u_\pm(t,k)\right|^2
=\lim_{s\rightarrow 0}\frac{k^3}{2\pi^2}\frac{\left|\yasy(s)\right|^2}{M(s)}\nonumber\\
&=\frac{G\hbar\, H_\ast^2}{2\pi}\left(1+2\epsilon_{H\ast}\left(1-\Gamma_E-\log\left(\frac{2k}{k_\ast}\right)\right)\right.\nonumber\\
&\left.\qquad+2\sigma_\ast\left(1-\frac{2(1\pm \pi \gamma)}{\gamma^2+1}\right)\right)\,+\,\mathcal{O}(\epsilon^2),&
\end{align}
where $\Gamma_E\approx0.577$ is Euler's constant. From this expression, we can read, at first order in the slow-roll parameters, the amplitude at the pivot mode $k_\ast$,
\begin{align}\label{Amplitude}
A_{\ast\pm}&\equiv\Delta^2_\pm(k_\ast)
=\frac{G_N\hbar H_\ast^2}{2\pi}\left(1+2\epsilon_{H\ast}\left(1-\Gamma_E-\log 2\right)\right.\nonumber\\
&\left.\qquad\qquad\quad\quad+2\sigma_\ast\left(1-\frac{2(1\pm \pi \gamma)}{\gamma^2+1}\right)\right)\,,
\end{align}
and the tensor tilt at the pivot mode,
\begin{equation}
n_{T\ast}\equiv \left.\frac{d\log(\Delta^2_\pm(k))}{d\log(k)}\right|_{k_\ast} =-2\epsilon_{H\ast}\,.
\label{tilt}
\end{equation}
While the amplitude depends on the circular polarization of the modes, at this order in the slow-roll expansion, the tilt does not. Finally, the polarization of the mode $k_\ast$ is
\begin{equation}\label{Polarization}
\Pi_\ast\equiv \frac{A_{\ast+}-A_{\ast-}}{A_{\ast+}+A_{\ast-}}=-4\pi\sigma_\ast\frac{\gamma}{\gamma^2+1}.
\end{equation}
The polarization is non-vanishing for $\sigma_\ast\neq0$ and depends on the value of the Barbero-Immirzi parameter $\gamma$, see Fig.~\ref{figure-gamma}. 

The expressions for the tilt $n_{T\ast}$ and the polarization $\Pi_\ast$ (\ref{tilt},\ref{Polarization}) are in agreement with the ones found in a more general case in \cite{SatohSoda,Satoh2010} where a different approximation for the mode equation is used. In particular the $\log(s)$ expansion \eqref{logsH} allows us to determine the late time limit \eqref{Power Dimless} in which the power spectrum is explicitly time independent.

\section{Discussion}\label{Section: Discussion measurements}

\begin{figure*}[t]
\includegraphics[height=0.30\textwidth]{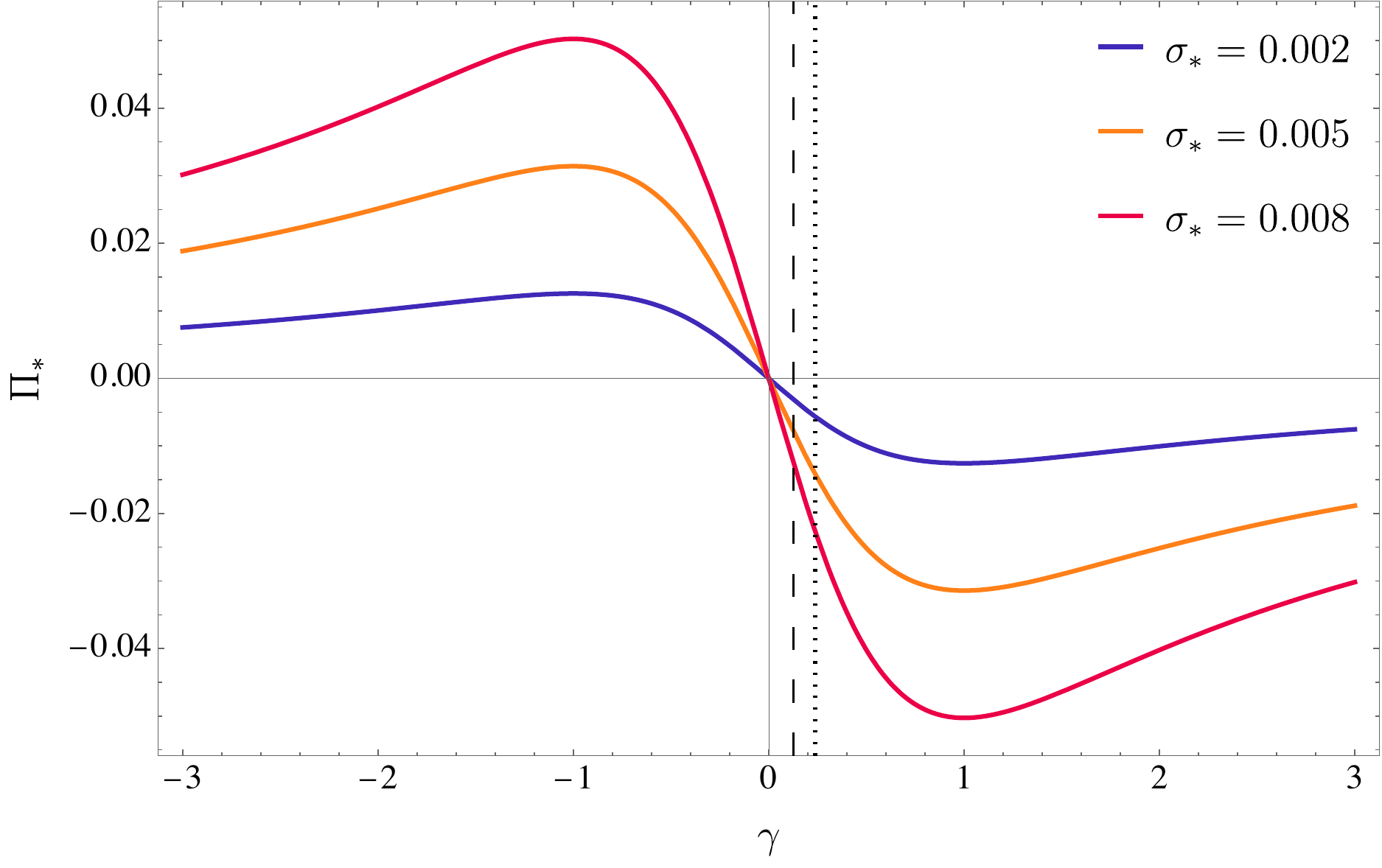}\hspace{2em}
\includegraphics[height=0.30\textwidth]{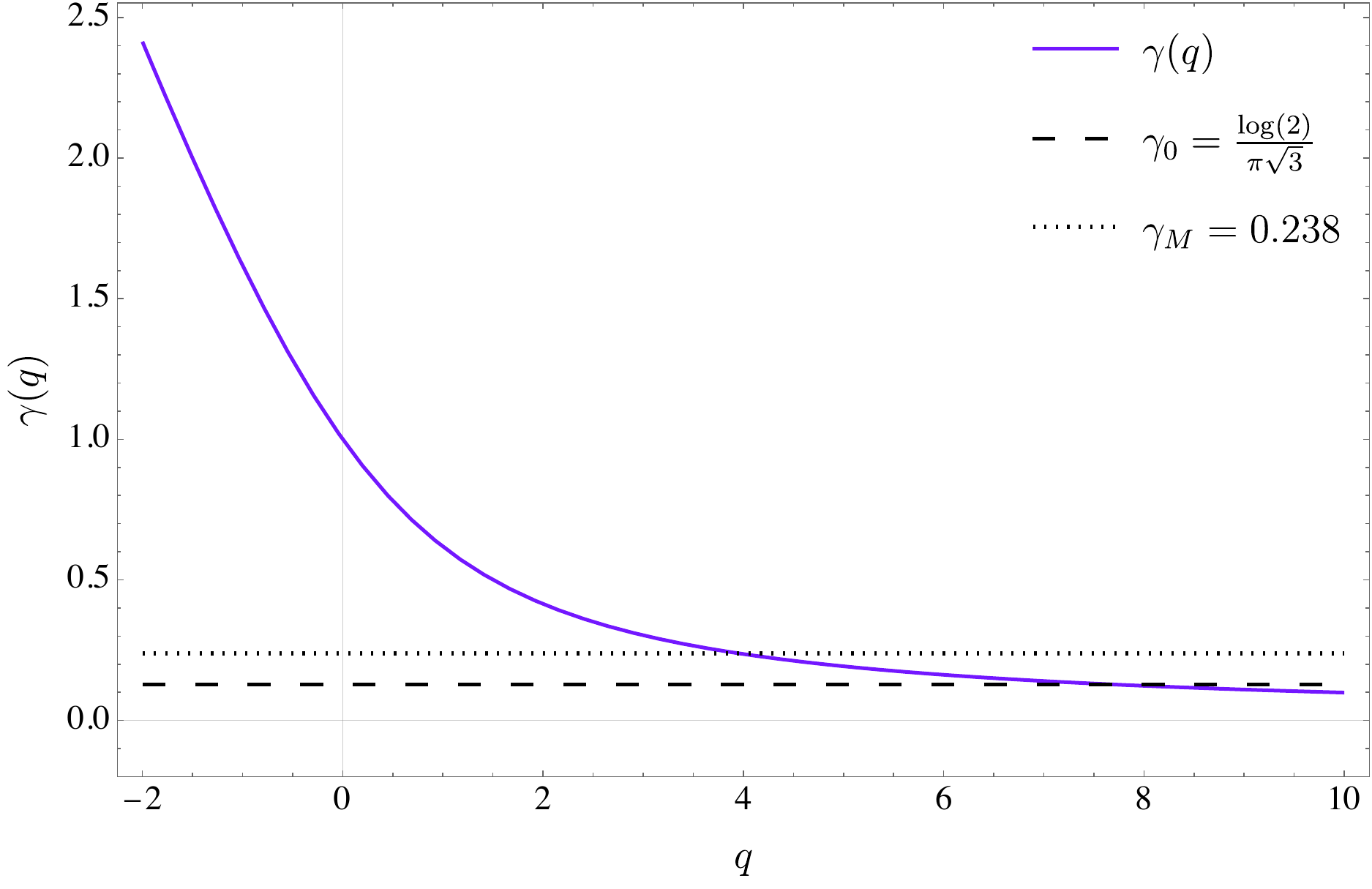}
\caption{(\emph{Left.}) Dependence of the polarization of primordial gravitational waves at the pivot mode $k_\ast$ on the Barbero-Immirzi parameter for the $\gamma$-dual action \eqref{Seff gDual}, assuming different values of the slow-roll parameter $\sigma_\ast$, \eqref{sigma_def}.  
(\emph{Right.}) Value of $\gamma$ determined by a possible measurement of the cosmological observable $q$, defined in terms of the tensor polarization, tensor tilt and tensor-to-scalar ratio (\ref{gamma-q}). For reference we report the values $\gamma_0\simeq 0.127$ (dashed line)  and 
$\gamma_M\simeq 0.238$ (dotted line) considered in \cite{AABaezCorichi} and \cite{Meissner2004}.
}
\label{figure-gamma}
\end{figure*}

In this paper we have introduced a new condition on parity-violating effective actions, the notion of $\gamma$-duality, that is motivated by the specific dependence of the spinfoam dynamics $W_\gamma$ on this parameter. While at present there is no top-down derivation of an effective field theory from the full non-perturbative spinfoam dynamics of loop quantum gravity, the $\gamma$-dual effective action \eqref{Seff gDual} encodes in a crucial way an exact property of the non-perturbative theory: spacetime diffeomorphisms $\mathrm{Diff}_0(\mathcal{M})$ that are connected to the identity are an exact gauge symmetry, while the invariance under large diffeomorphisms such as orientation reversals $\mathcal{R}$ is broken in a way that is controlled by the single parameter $\gamma$. 

The possibility of determining the Barbero-Immirzi parameter $\gamma$ from observations was first considered in \cite{ContaldiMagueijoSmolin} as a measure of parity violation in primordial gravitational waves, and then in \cite{Perez:2005pm,Freidel:2005sn} as a parity-violating coupling of fermions to gravity. This condition of $\gamma$-duality relates the coupling constants in the effective action and therefore allows us to probe the dependence of observables on the parameter $\gamma$ at the level of higher-curvature terms. 

\medskip

As an effective theory for cosmological inflation, the $\gamma$-dual action \eqref{Seff gDual} predicts a polarization for primordial gravitational waves given by the relation \eqref{Polarization},
\begin{equation}\label{Pi-sigma-gamma}
\Pi\equiv \frac{A_{+}-A_{-}}{A_{+}+A_{-}}=-4\pi\sigma_\ast\frac{\gamma}{\gamma^2+1}.
\end{equation}
The difference in amplitude between the two circular polarizations is a direct consequence of the parity-violating nature of the action \cite{LueWangKamionkowski,AlexanderYunes,SatohSoda,Satoh2010}. Besides the dependence on $\gamma$, we note that the polarization $\Pi$ depends also on the value of the slow-roll parameter $\sigma_\ast$ at the pivot mode $k_\ast$, which measures the shape of the coupling function $f(\phi)$ as defined in (\ref{sigma_def}), (see Fig.~\ref{figure-gamma}, Left). Therefore, determining $\gamma$ requires an independent measurement of other observables besides the tensor polarization $\Pi$. A similar dependence on additional parameters besides $\gamma$ appears in the analysis of parity-violating observables in the coupling of fermions to gravity \cite{Perez:2005pm,Freidel:2005sn}.

\medskip

The scalar power spectrum in an effective theory of the class considered here was derived in \cite{SatohSoda,Satoh2010}, assuming general coupling functions  $f_{GB}(\phi)$ and $f_{CS}(\phi)$ to the Gauss-Bonnet and Chern-Simons terms. Expressing the result of \cite{SatohSoda,Satoh2010} in terms the coupling functions of the $\gamma$-dual action \eqref{Seff gDual}, we find that the tensor-to-scalar ratio at the pivot mode $k_\ast$ is
\begin{equation}\label{r scalar-to-tensor}
r = 16 \,\epsilon_{H\ast}+32\gbc\sigma_\ast\,,
\end{equation}
which depends on the slow-roll parameters $\epsilon_{H\ast}$ and $\sigma_\ast$ at the pivot mode, besides $\gamma$. Moreover, the tensor tilt is given by the expression (\ref{tilt}),
\begin{equation}
n_{T}=-2\epsilon_{H\ast}\,.
\end{equation}
Combining the primordial cosmological observables $\Pi$, $r$ and $n_T$, we can extract a relation that depends only on the Barbero-Immirzi parameter $\gamma$,
\begin{equation}\label{1/gamma-gamma}
\frac{1}{\gamma}-\gamma=\frac{\pi}{8}\frac{8n_{T}+r}{\Pi}.
\end{equation}
The relation \eqref{1/gamma-gamma} can be solved for  positive value of the Barbero-Immirzi parameter in terms of the ratio $q$,  (see Fig.~\ref{figure-gamma}, Right):
\begin{equation}\label{gamma-q}
\gamma\,=\,\sqrt{1+\left(\frac{q}{2}\right)^2\,}-\frac{q}{2}\,,\quad\textrm{with}\;\; q\equiv\frac{\pi}{8}\frac{8n_{T}+r}{\Pi}.
\end{equation}
In this way, a measurement of the polarization and tilt of primordial gravitational waves, together with their relative amplitude to the scalar modes, can provide a measurement of the Barbero-Immirzi parameter and, therefore, of the scale of the discreteness \eqref{area spectrum} in loop quantum gravity.

\medskip

\comment{We can compare the result \eqref{1/gamma-gamma} to the case of two general coupling functions $f_{GB}(\phi)$ and $f_{CS}(\phi)$ as defined in \eqref{L_GB},\eqref{L_CS}. In this case, the cosmological observable $q=\frac{\pi}{8}(8n_T+r)/\Pi$ will depend on the derivatives of these functions as discussed in \cite{SatohSoda, Satoh2010}. If some fundamental principle implies that the ratio between the two functions is constant, $f_{GB}(\phi)/f_{CS}(\phi)=\mathrm{const}$, then the cosmological observable $q$ will depend only on this constant. By itself, this property is not unique to $\gamma$-duality. The requirement of $\gamma$-duality imposes that the ratio is constant and given by a specific expression in terms of the Barbero-Immirzi parameter $\gamma$. }

\medskip

The relation \eqref{gamma-q} arises from the analysis of slow-roll inflation with the $\gamma$-dual effective action \eqref{Seff gDual}. We expect a similar relation to arise when one starts from a more general parity-non-violating action that includes also $K$-inflation terms and is then duality-rotated to a $\gamma$-dual action. It is then interesting to investigate the possibility of determining $\gamma$ in such theories using future cosmological observations of the polarization $\Pi$ that appears in the denominator in \eqref{gamma-q}, and of the consistency relation $r+8n_{T}$ that appears in the numerator:

(a) \emph{Polarization}. There are prospects of measurements of the polarization $\Pi$ in the near future with Cosmic Microwave Background (CMB) polarization experiments, including BICEP2/Keck \cite{BICEP2Keck2018}, SPT \cite{SPT2020}, SPIDER \cite{SPIDER2018}, LiteBIRD \cite{LiteBIRD}, and CMB-S4 \cite{CMBS4PGW} experiments. For a comprehensive review on the main challenges and strategies for extracting the contribution from primordial gravitational waves to the $B$-modes in the CMB, see \cite{KamionkowskiKovetzReview}. While scalar perturbations can only imprint (even) E-polarization modes on the CMB, tensor perturbations can induce both (even) $E$ and (odd) $B$-polarization modes 
\cite{KamionkowskiKosowskiStebbins, SeljakZaldarriaga}. Furthermore, a measurement of non-zero two-point correlations between the temperature and $B$-modes of the CMB, $C_l^{TB}\neq 0$, or between the $E$ and $B$-modes, $C_l^{EB}\neq 0$, would be a definite indication of a parity violating phenomenon \cite{KamionkowskiKosowskiStebbins}. As shown in \cite{ContaldiMagueijoSmolin}, the contribution from tensor modes to $C_l^{EB}$ and $C_l^{TE}$ go respectively as $\Delta^2_+(k)-\Delta^2_-(k)$ (polarization) and $\Delta^2_+(k)+\Delta^2_-(k)$ (average amplitude), convoluted with the corresponding radiation transfer functions; and therefore, their measurement could be used to estimate the tensor polarization. For instance, in \cite{FerteGrain} one can find an analysis of the constraints on tensor polarization from the $EB$ and $TB$ correlations and the prospects for satellite-like missions for models with parity violation and a polarization $\Pi$.  

(b) \emph{Consistency relations}. CMB-S4 \cite{CMBS4PGW} experiments are expected to measure, or put stronger bounds on, the tensor-to-scalar ratio $r$ in the near future. Together with a measurement of the tensor tilt $n_T$, these observations will provide a probe of the consistency relation $r=-8n_{T}$, a property of single-field slow-roll inflation in pure General Relativity. The effective theory described by the action (\ref{Seff gDual}) violates the consistency relations and the quantity $r+8n_{T}$ appears in the numerator in \eqref{1/gamma-gamma}. While, in principle, we can have $|r+8n_{T}|\ll 1$ and $|\Pi|\ll1$, which would make observations difficult, the value of $\gamma$ depends only on their ratio. For $|r+8n_{T}|\ll |\Pi|$, we find a Barbero-Immirzi parameter $\gamma\approx 1$, while for $|r+8n_{T}|\gg |\Pi|$ we have $\gamma\ll 1$.

\medskip

\comment{In the analysis of Sec.~\ref{sec: Inflation and PGW}, we assumed that the slow-roll inflationary phase is sufficiently long and the power spectrum of perturbations is determined by the Bunch-Davies-like vacuum state \eqref{yad}. In loop quantum cosmology \cite{MB:SingularityLQC, AA-MB-JL:LQC-Math,AA-PS:LQC, IA-PS:LQC, IA-EWEea}, the pre-inflationary phase results in a squeezed state and deviations from an almost scale-invariant power spectrum. In particular, in a ``just-enough inflation'' scenario with the slow-roll phase lasting about $60$ e-foldings, primordial features and power suppression are expected to appear in the power spectrum at large scales \cite{Agullo:2012sh,Fernandez-Mendez:2012poe,Fernandez-Mendez:2013jqa,Agullo:2013ai,Agullo:2015tca,deBlas:2016puz,AshtekarGupt2016, AshtekarGupt2017, Navascues2020,Ashtekar:2020gec,Navascues:2021mxq,Martin-Benito:2023nky,ElizagaNavascues:2023xah,MenaMarugan:2024vyy,Garay:2024afl,Montese:2024ypz}. It would be interesting to extend the analysis presented here to squeezed vacua that take these effects into account.}

\medskip

Here we focused exclusively on the primordial power spectrum. It would be interesting to extend this analysis of bounds on the Barbero-Immirzi parameter via $\gamma$-dual effective actions for inflation, to other cosmological observables such as parity-violating features in primordial non-Gaussianities \cite{Bartolo:2014hwa,Bartolo:2015dga,Bartolo:2017szm,Nojiri:2019nar,Bordin:2020eui,Alexander:2022cow,Gong:2023kpe,Jenks:2023pmk,Creque-Sarbinowski:2023wmb}, and their possible imprint in the large scale structure of the Universe  \cite{DodelsonStebbins, Donghui-Shear, DonghuiDimastrogiovanni-LSS} and galaxy shapes \cite{GalaxyShapesPhilcoxSA}.\\

\medskip

\emph{Acknowledgments.}
We thank Pietro Don\`a, Juan Margalef-Bentabol, Djordje Minic, Javier Olmedo, and Marc Schneider for useful discussions. This work was made possible through the support of the Grant ID 62312 and Grant ID 63683 from the John Templeton Foundation (JTF), as part of the project ``The Quantum Information Structure of Spacetime'' (QISS) and the WOST project (\href{https://withoutspacetime.org}{\mbox{withoutspacetime.org}}). The opinions expressed in this work are those of the authors and do not necessarily reflect the views of the John Templeton Foundation. E.B.~acknowledges support from the National Science Foundation, Grants No. PHY-2207851 and PHY-2513194. M.R.R. acknowledges support from the U.S. Department of Energy (DOE) under award number~DE-SC00019066.

\providecommand{\noopsort}[1]{}\providecommand{\singleletter}[1]{#1}%
%

%

\end{document}